\documentclass[sigconf,nonacm]{acmart}
\usepackage{amsmath}
\usepackage{graphicx}
\usepackage{here}
\PassOptionsToPackage{hyphens}{url}
\usepackage{hyperref}
\usepackage{url}
\usepackage{textcomp}
\usepackage{multicol}
\usepackage{subcaption}
\usepackage{comment}
\usepackage{booktabs,dcolumn}
\usepackage{multirow}
\usepackage{siunitx}
\usepackage{fancyhdr}
\renewenvironment{quote}[1][0.04\linewidth]
  {\list{}{\leftmargin=#1\rightmargin=#1}\item\relax}{\endlist}

\makeatletter
\def\Hline{
  \noalign{\ifnum0=`}\fi\hrule \@height 4.\arrayrulewidth \futurelet
   \reserved@a\@xhline}
\makeatother
\AtBeginDocument{%
  }
\setcopyright{acmlicensed}
\copyrightyear{2018}
\acmYear{2018}
\acmDOI{XXXXXXX.XXXXXXX}
\acmJournal{JACM}
\acmVolume{37}
\acmNumber{4}
\acmArticle{111}
\acmMonth{8}
\usepackage[dvipsnames]{xcolor}
\usepackage{subcaption}
\newcommand{\tabitem}{~~\llap{\textbullet}~~}
\definecolor{teal}{HTML}{0072B2}
\definecolor{redorange}{HTML}{D55E00}
\usepackage[table]{xcolor}
\usepackage{enumitem}
\begin{document}
\title{Understanding Reader Perception Shifts upon Disclosure of AI Authorship}
\author{Hiroki Nakano}
\authornote{Equal contribution.}
\email{nakahiro@iis-lab.org}
\affiliation{%
  \institution{IIS Lab, The University of Tokyo}
  \state{Tokyo}
  \country{Japan}
}
\author{Jo Takezawa}
\authornotemark[1]
\email{takezawa@iis-lab.org}
\affiliation{%
  \institution{IIS Lab, The University of Tokyo}
  \state{Tokyo}
  \country{Japan}
}
\author{Fabrice Matulic}
\email{fmatulic@preferred.jp}
\affiliation{%
  \institution{Preferred Networks Inc.}
  \state{Tokyo}
  \country{Japan}
}
\author{Chi-Lan Yang}
\email{chilan.yang@iii.u-tokyo.ac.jp}
\affiliation{%
  \institution{IIS Lab, The University of Tokyo}
  \state{Tokyo}
  \country{Japan}
}
\author{Koji Yatani}
\email{koji@iis-lab.org}
\affiliation{%
  \institution{IIS Lab, The University of Tokyo}
  \state{Tokyo}
  \country{Japan}
}
\begin{abstract}
As AI writing support becomes ubiquitous, the question of how disclosing its use affects reader perception remains critical and underexplored. We conducted a controlled study with 261 participants to examine how disclosing varying levels of AI involvement shifts perceptions of the author across six distinct communicative acts. Our analysis of 990 evaluations reveals that disclosure generally erodes perceived trustworthiness, caring, competence, and likability, with the most precipitous declines observed in social and interpersonal writing. A thematic analysis of participant feedback attributes these negative shifts to a perceived loss of human sincerity, diminished authorial effort, and the contextual inappropriateness of AI. Notably, however, we find that higher AI literacy mitigates these negative perceptions, leading to greater tolerance or even appreciation for AI assistance. Our results highlight the nuanced social dynamics of AI-mediated authorship and inform design implications for transparent, context-sensitive writing systems that better preserve trust and authenticity.
\end{abstract}

\thanks{Accepted at 31st International Conference on Intelligent User Interfaces (IUI '26).}

\maketitle
\section{Introduction}
Artificial intelligence (AI) systems have reached a level of writing fluency comparable to human authors leading to their widespread adoption across professional, creative, and interpersonal contexts.
While these tools can enhance clarity and expression, they also introduce significant social frictions.
Readers often perceive AI-generated texts as less empathetic or authentic~\cite{rubin2025comparing}, and disclosing AI authorship can degrade perceptions of authenticity and trust~\cite{jakesch2019ai, liu2022will, schilke2025transparency}.
This tension arises because writing is more than mere information transfer. It is a social act that conveys an author’s intention, effort, and emotion to readers.
When an AI actively contributes to the writing process, this fundamental social meaning of authorship becomes ambiguous, complicating how a text and its author are perceived.

While previous work in AI-mediated communication has outlined these ethical and relational challenges~\cite{hancock2020ai}, empirical studies have so far examined reader perceptions with limited granularity or control over AI authorship.
Much of the existing research compares purely manual writing to AI-assisted text in specific, narrow contexts, often treating writing as a monolithic activity~\cite{jain2024revealing, liu2022will, schilke2025transparency}.
This overlooks the nuanced nature of writing, which serves a wide spectrum of communicative goals.
As articulated in frameworks like Berge et al.'s \textit{Wheel of Writing}, writing can be used to persuade, to interact socially, to reflect internally, or to imagine creatively~\cite{berge2016wheel}.
The social expectations for an apology letter, for example, are vastly different from those for a technical manual.
As AI writing support becomes deeply embedded in these varied contexts, the boundary between helpful assistance and core authorship blurs.
Furthermore, readers' attitudes toward writing co-authored by AI may vary depending on their familiarity with the technology. Previous studies have indicated that individuals who are more knowledgeable about AI tend to view its use as a pragmatic choice rather than a lack of competence~\cite{mahmud2024decoding}.
This raises critical, unanswered questions: How do readers' perceptions shift based on the \textit{degree} of AI authorship across these different communication purposes? Furthermore, how do individual factors, such as readers' AI literacy, moderate these perceptions?
Addressing these questions is essential for designing AI authoring tools that are not only effective but also socially acceptable and trustworthy.

To address this research gap, we conducted a controlled user study with 261 participants that examines how readers’ perceptions of a fictitious author change upon the disclosure of AI authorship across six distinct acts of writing~\cite{berge2016wheel}.
Our analysis revealed consistent negative shifts in perceived trustworthiness, caring, competence, and likability, especially in emotional and interpersonal writing.
However, we also found that participants with higher AI literacy exhibited greater tolerance toward, and even appreciation of, AI authorship.
Our results highlight the complex social dynamics of AI authorship and inform the design of transparent, context-sensitive writing tools that preserve authenticity and trust.
We position AI writing assistance not merely as a technical problem of automation, but as a relational design challenge.
We argue that future interfaces must balance AI capabilities with the need to communicate human authorship, social context, and the perceived effort invested by the author.

In summary, we make the following contributions to research on AI-mediated writing and human–AI interaction:

\begin{itemize}
    \item \textbf{Granular empirical evidence on perception shifts after AI authorship disclosure}: We present findings from a controlled study (N=261) that systematically investigates how reader impressions of an author change after disclosing varied levels of AI authorship across six different acts of writing. Our results reveal consistent negative perception shifts in trust, caring, competence, and likability, with stronger effects observed in interpersonal contexts.
    \item \textbf{Insights into the social dynamics of AI-mediated writing}: Our quantitative and qualitative analyses show that perceptions are shaped not only by the degree of AI authorship but also by perceived human effort, emotional context, and readers’ AI literacy. These factors provide a foundation for designing more socially aware AI authoring systems.
    \item \textbf{Design implications for future AI-mediated writing}: Based on our findings, we derive four design implications for future systems: 1) context-sensitive transparency that explains the purpose of AI authorship; 2) interfaces that preserve and communicate human effort and agency; 3) writing support adaptive to emotional and social contexts; and 4) reflective interfaces that foster AI literacy and calibrated trust.
\end{itemize}

\section{Related Work}

We situate our work within two primary streams of research: the application of AI across the diverse functional purposes of writing, and the growing body of literature on how readers perceive and react to the disclosure of AI authorship in communication.

\subsection{AI-assisted Writing for Diverse Communicative Purposes}
Writing serves a wide spectrum of goals, from the interpersonal to the informational. \citet{berge2016wheel} outline six fundamental purposes: persuading audiences, interacting with others, reflecting on experiences, describing facts, exploring ideas, and creating imaginative worlds. Modern AI-assisted writing tools are increasingly used across all of these distinct cognitive and social functions.

Prior research has explored how AI tools support a range of writing goals. 
Mirowski et al. demonstrated that AI can be integrated into writers' workflows to enhance creativity in screen- and playwriting \cite{mirowski2023co}.
AI has also been leveraged in contexts where writing serves an interpersonal or social function.
For instance, systems that suggest, generate, or refine email responses can help users convey positivity~\cite{mieczkowski2021ai}, politeness, and professionalism in workplace interactions~\cite{Miura2025Understanding}, while AI-assisted dating profiles can enhance self-presentation~\cite{Barkallah2025Transparent}.
These applications demonstrate that the role of AI extends far beyond grammatical correction to shaping a writer's tone, style, and communicative intent, thereby aiding in both persuasion and social engagement~\cite{hancock2020ai}.

This expanding role, however, introduces social friction, particularly when AI's contribution blurs the boundaries of authorship. Readers often question the authenticity and credibility of AI-generated content. 
For instance, ~\citeauthor{li2024does} found that revealing the role of AI in essay writing lowered quality ratings from readers~\cite{li2024does}.
Similarly, research on AI-mediated communication showed that online profiles perceived as AI-generated are less trusted~\cite{jakesch2019ai}.
These concerns are particularly salient when writing serves a social purpose, such as persuasion or interaction, where impressions of the author are heavily shaped by perceived authenticity. Recent studies on disclosure dynamics confirm this tension: while AI assistance can improve a text's fluency, revealing its use can lead readers to judge the author as less sincere or capable~\cite{hwang202580, draxler2024ai}.

AI-assisted writing thus presents a fundamental trade-off. On one hand, it empowers individuals, especially those with language barriers or weaker writing skills, to communicate more effectively~\cite{ningrum2023chatgpt, noy2023experimental}. On the other, the social meaning of authorship complicates its adoption. When readers know or suspect AI involvement, they may perceive the writing---and by extension, the author---as less authentic or trustworthy~\cite{jakesch2019ai}. 
Understanding how these perceptions vary across different communicative purposes is therefore critical for designing AI tools that successfully balance expressive power with the need to maintain credibility.

\subsection{Reader Perceptions of AI Disclosure}

Research on AI disclosure highlights its paradoxical effect on interpersonal trust.
Several studies show that when readers learn AI contributed to a text, their trust in the human author declines even if the quality of the text is unchanged~\cite{jakesch2019ai, lermann2024effects, liu2022will, proksch2024impact, schilke2025transparency}. This ``transparency penalty'' challenges the conventional assumption that openness builds trust~\cite{sah2018conflict}.
The magnitude of this trust erosion is not uniform and appears to be moderated by individual differences. For instance, \citet{schilke2025transparency} found that people with more positive attitudes toward AI show a smaller decline in trust upon disclosure. Furthermore, a user's AI literacy plays a significant role in shaping their perceptions~\cite{mahmud2024decoding, schiavo2024comprehension}. AI-literate individuals tend to view its use as pragmatic, whereas less knowledgeable readers may interpret it as a sign of incompetence or laziness~\cite{mahmud2024decoding, schiavo2024comprehension}.

Complicating this landscape is the fact that people are increasingly unable to distinguish AI-generated text from human writing without explicit disclosure~\cite{majovsky2024perfect, weber2023testing}, a finding that holds true even for those knowledgeable about large language models (LLMs)~\cite{dugan2023real, hakam2024human, porter2024ai}. This creates a risk of misattribution, where readers may credit AI-generated text to a human, only to have their perception negatively altered if AI involvement is later revealed. Despite the importance of this phenomenon, little research has systematically examined how varying \textit{degrees} of AI involvement influence reader perception across different writing contexts.

Beyond trust, research also reveals mixed attitudes toward the social qualities of AI-generated text. 
Some studies find that AI generation in communicative scenarios can trigger an ``Uncanny Valley effect'', eliciting discomfort~\cite{radivojevic2024human, yin2024ai}.
Others indicate that AI is perceived as less caring, warm, or empathetic than humans, particularly in sensitive contexts like health communication~\cite{rubin2025comparing, wang2025chatgpt}. 
Conversely, other studies show that AI-generated content can be rated as equal or even superior to human-authored text, especially when qualities like efficiency and accuracy are valued~\cite{ayers2023comparing, ovsyannikova2025third}.

One factor that may explain these divergent findings is people's prior experience and familiarity with AI. Regular users may view AI authorship as beneficial~\cite{mahmud2024decoding}, while less experienced individuals might perceive it as cold, inauthentic or inadequate~\cite{li2024does}. Together, these findings suggest that AI disclosure has multifaceted effects, influencing not only trust in the author but also perceptions of warmth, competence, and empathy.

\subsection{Our Approach}

Building on this body of work, our study introduces a more granular approach to understanding the effects of AI authorship disclosure. Unlike research focused on single domains or simple binary comparisons (human vs. AI-assisted), we systematically investigate these effects across six distinct communicative acts while also varying the disclosed \textit{degree} of AI authorship. This multi-faceted design allows us to model how reader perceptions shift as AI's role moves along a continuum from minor assistant to primary author, providing a more nuanced understanding of the social acceptability of human-AI co-authorship.

\section{Method}

To investigate how the disclosure of AI assistance influences reader perception, we conducted a repeated-measures online study. Participants rated their impressions of a text before and after being informed that a specific percentage of the content was generated or edited by AI. Unlike prior work, our primary objective was to examine how perceptions shift across systematically varying \textit{levels} of disclosed AI contribution. To achieve this, we used a deception-based design: while all texts were entirely AI-generated (with minor human proofreading), participants were told that only specific portions were created or altered by AI. This approach enabled precise control over the perceived degree of AI involvement.

Our study addressed the following research questions:

\begin{enumerate}[label=\textbf{RQ\arabic*}, ref=RQ\arabic*]
	\item How does the communicative purpose (i.e., the act of writing) of a text shape reader perceptions upon the disclosure of AI assistance? \label{rq:act-of-writing}
	\item How does a reader's AI literacy moderate these perception shifts? \label{rq:ai-literacy}
	\item What qualitative reasons and underlying themes explain the changes in perception following AI disclosure? \label{rq:perception}
\end{enumerate}

The following study protocol was approved by the Institutional Review Board of the first author's university.

\subsection{Experiment Design}
We grounded our experimental design in the six fundamental acts of writing defined by Berge et al.~\cite{berge2016wheel}: 
\textit{Convince}, \textit{Interact}, \textit{Reflect}, \textit{Describe}, \textit{Explore}, and \textit{Imagine}.
As detailed in Table~\ref{tab:writing_purposes}, these acts serve distinct communicative purposes. \textit{Convince}, \textit{Interact}, and \textit{Reflect} are primarily person-oriented, focusing on others, social relationships, and the self, respectively. In contrast, \textit{Describe}, \textit{Explore}, and \textit{Imagine} are object-oriented, centering on the organization, development, and creation of knowledge.

\begin{table*}[tb]
  \small
  \centering
  \caption{The six acts of writing and their communicative purposes, adapted from Berge et al.~\cite{berge2016wheel}.}
  \begin{tabular}{p{0.4\textwidth}l}
    \toprule
    \textbf{Act} & \textbf{Purpose} \\
     \hline
    \textbf{To convince}, express meanings, argue, discuss, and recommend & Persuasion \\ 
    \textbf{To interact}, cooperate & Exchange of information, being in dialogue \\
    \textbf{To reflect} upon own experiences, thoughts, feelings, and work & Identify information, self-evaluation and meta-communication \\
    \textbf{To describe}, organize and structure content & Knowledge organization and storing \\
    \textbf{To explore}, investigate, compare, analyze, discuss, interpret, explain, and reason & Knowledge development \\
    \textbf{To imagine}, narrate, create, and theorize & Creation of textual worlds \\
    \bottomrule
  \end{tabular}
  \label{tab:writing_purposes}
\end{table*}

To create our experimental stimuli, we generated three scenarios for each act, yielding 18 unique texts (Table~\ref{tab:scenarios}).
We created 3--5 initial scenarios per act using GPT-4o\footnote{\url{https://chatgpt.com}} by providing the model with the definition of the act. We then selected the three scenarios we found the most suitable and prompted GPT-4o to generate sentences based on their descriptions.
As the study targeted Japanese participants, all texts were presented in Japanese. 
Pilot testing revealed that sentences generated directly in Japanese by GPT-4o lacked natural fluency. Therefore, we first generated the text in English and then translated it into Japanese.
A native Japanese-speaking author reviewed all translated texts, making minor corrections to ensure linguistic quality and naturalness. The final texts averaged 20 sentences and 662 Japanese characters in length. The full English and Japanese scripts are provided as supplementary material.

\begin{table*}[t]
  \small
  \centering
  \caption{The 18 scenarios used as experimental stimuli, organized by the six acts of writing.}
  \begin{tabular}{p{3cm}l}
    \toprule
    \textbf{Act of writing} & \textbf{Scenarios} \\
    \midrule
    \textbf{To convince} & 
    \begin{tabular}{@{}l@{}}
      \tabitem A persuasive comment on a news article \\ 
      \tabitem A political campaign speech \\
      \tabitem An apology letter to resolve a conflict \\
    \end{tabular}
    \\ \hline
    \textbf{To interact} & 
    \begin{tabular}{@{}l@{}}
      \tabitem An email to a friend after a long time \\
      \tabitem A gratitude letter \\
      \tabitem An encouragement letter to a friend in the hospital \\
    \end{tabular}
    \\ \hline
    \textbf{To reflect} & 
    \begin{tabular}{@{}l@{}}
      \tabitem A personal diary entry \\
      \tabitem A reflection on one's career and life path \\
      \tabitem A reflection on human growth and maturity \\
    \end{tabular}
    \\ \hline
    \textbf{To describe} & 
    \begin{tabular}{@{}l@{}}
      \tabitem An encyclopedia entry about an animal \\
      \tabitem A news article reporting a natural disaster \\
      \tabitem A tutorial on operating new equipment \\
    \end{tabular}
    \\ \hline
    \textbf{To explore} & 
    \begin{tabular}{@{}l@{}}
      \tabitem An academic essay disputing a scientific theory \\
      \tabitem A blog post discussing the message of a book \\
      \tabitem A report analyzing the impact of a new policy \\
    \end{tabular}
    \\ \hline
    \textbf{To imagine} & 
    \begin{tabular}{@{}l@{}}
      \tabitem A passage from a novel \\
      \tabitem A passage from a science fiction story \\
      \tabitem A prose poem \\
    \end{tabular}
    \\ \hline
    \bottomrule
  \end{tabular}
  \label{tab:scenarios}
\end{table*}

To manipulate the perceived amount of AI involvement, we simulated various levels of AI authorship. For each text, a random proportion of sentences was labeled as generated or edited by AI (in reality, all texts were entirely AI-authored).
The distribution of these AI-labeled sentences was randomized using a seed unique to each condition.
The disclosed proportion varied from 0\% to 100\% in 10\% increments across conditions, corresponding to approximately two sentences per increment (Figure~\ref{fig:experiment_workflow}). 
This design allowed us to model situations where a reader discovers that some portions of a text, from minor local edits to the entire document, were created with AI assistance.

\subsection{Procedure}
The experimental workflow consisted of three main parts for each text evaluated by participants.

\subsubsection{Pre-disclosure Evaluation}
\begin{figure}[t]
  \centering
  \begin{subfigure}[t]{0.48\textwidth}
    \centering
    \includegraphics[width=\textwidth]{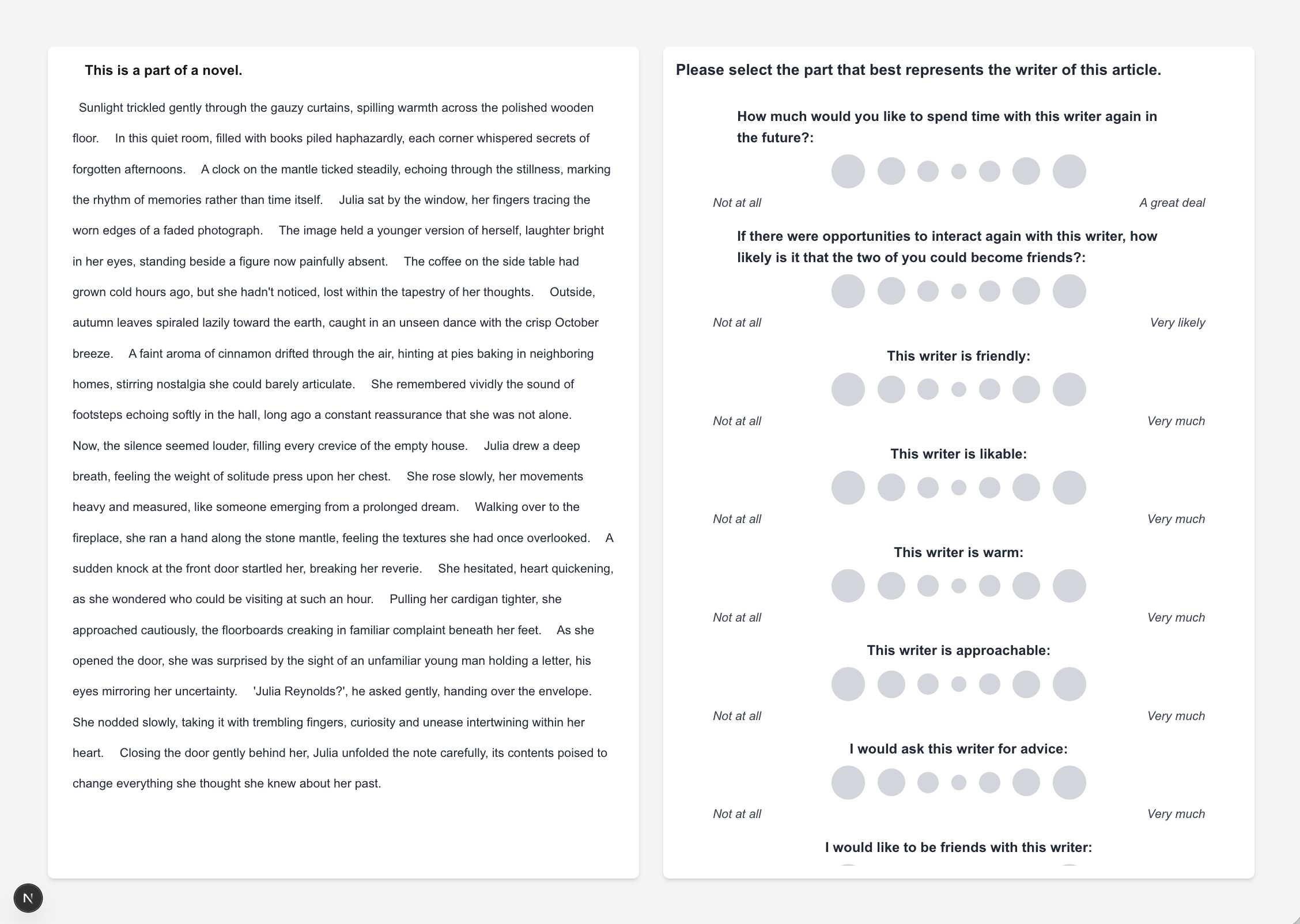}
    \caption{The interface shown during the pre-disclosure phase. The left panel displays the text and the right panel shows the questionnaire.}
    \label{fig:pre-rating}
  \end{subfigure}
  \hfill
  \begin{subfigure}[t]{0.48\textwidth}
    \centering
    \includegraphics[width=\textwidth]{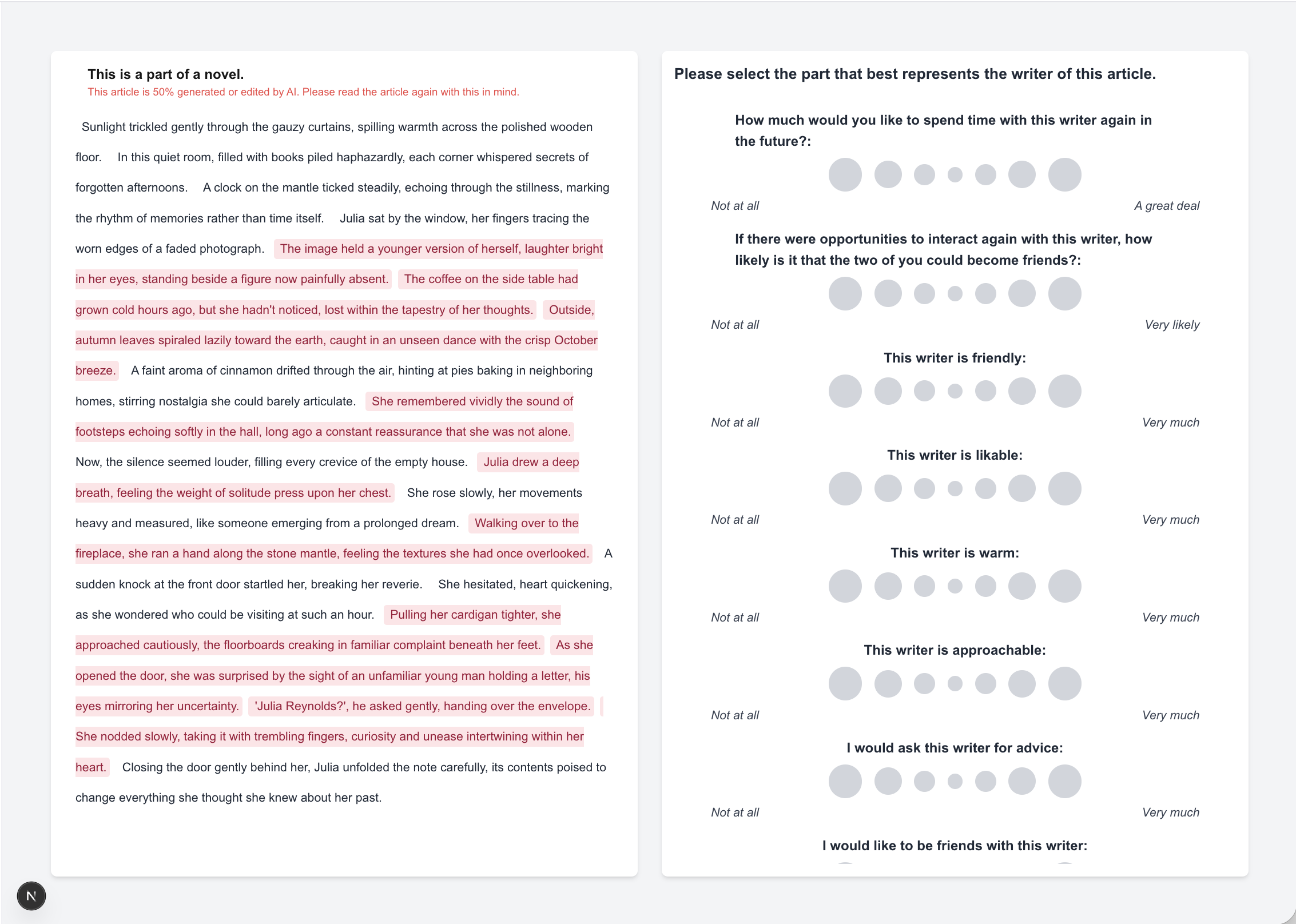}
    \caption{The interface shown during the post-disclosure phase. The text is displayed with highlighted sentences indicating content (supposedly) generated/edited by AI, alongside the same questionnaire.}
    \label{fig:post-rating}
  \end{subfigure}
  \caption{The user interfaces of the experiment: (a) pre-rating and (b) post-rating phases. The original interface was in Japanese and the text shown here is translated for presentation, but the layout and visual design are preserved.}
  \Description{A two-panel figure. Part (a) shows the pre-disclosure interface with a block of text on the left and a questionnaire on the right. Part (b) shows the post-disclosure interface, which is identical except that several sentences in the left text panel are highlighted in red to indicate AI generation.}
  \label{fig:experiment_workflow}
\end{figure}

After providing informed consent, participants were each assigned four of the 18 texts, representing different acts of writing, and were asked to read them carefully. For each text, they answered a questionnaire assessing their initial impressions of the (fictitious) author (Figure~\ref{fig:pre-rating}).
The questionnaire assessed five perception dimensions using 7-point Likert scales adapted from established instruments (27 statements in total).
\textbf{Trustworthiness} (6 statements), \textbf{Caring/Goodwill} (6 statements), and \textbf{Competence}  (6 statements) were adapted from McCroskey and Teven~\cite{mccroskey1999goodwill}, and responses ranged from -3 (very negative) to +3 (very positive).
\textbf{Likability} (7 statements) and the participant's \textbf{Desire for Future Interaction} (2 statements), were adapted from Reysen~\cite{reysen2005construction} and Sprecher~\cite{sprecher2021closeness}.
These were rated from -3 (\textit{Strongly disagree}) to 3 (\textit{Strongly agree}).
All statements are included in Tables~\ref{tab:author-impression} and~\ref{tab:literacy} in the Appendix.

The same set of questions was used for both pre- and post-disclosure evaluations, including reverse-coded items and attention checks to ensure data quality.

\subsubsection{Simulated AI Authorship Disclosure and Post-disclosure Evaluation}

After submitting their initial ratings for a text, participants were informed that ``XX \% of the text was generated or edited by AI''. The corresponding sentences were then highlighted in the document viewer, similar to how tools like HaLLMark visualize AI-produced content~\cite{hoque2024hallmark}.

Following this disclosure, participants answered the same five questions to capture their revised impressions. 
They were also asked to provide a free-form text response explaining the rationale for any changes in their ratings.

\subsubsection{Post-experimental Questionnaire}
After completing evaluations for all four texts, participants filled out a final AI literacy questionnaire to determine whether their familiarity and skills with AI tools might have influenced their perceptions.
To measure this competency, we adapted three relevant sub-constructs from the Meta AI Literacy Scale (MAILS)~\cite{carolus2023mails} that assess familiarity with and control over AI: ``Apply AI'', ``Detect AI'', and ``AI Persuasion Literacy''.
We excluded other MAILS sub-constructs deemed less relevant to the specific scenarios in this study.
``Apply AI'' measures an individual's ability to use AI concepts for problem-solving, reflecting proficiency and comfort with AI tools~\cite{ng2022using}.
``Detect AI'' assesses the ability to distinguish AI-enabled technologies from those without AI, a core component of understanding the technology~\cite{long2020ai, wang2023measuring}.
Finally, ``AI Persuasion Literacy'' refers to the ability to recognize and resist AI-driven influence on one's decisions.
Table~\ref{tab:literacy} lists the items for each aspect of AI literacy.

\subsection{Participants}
We recruited 264 participants from a Japanese crowdsourcing platform~\footnote{\url{https://crowdworks.jp/}}.
The task was initially described as ``a survey to evaluate the quality and impression of written messages'', concealing the true purpose of the study.
After the initial pre-disclosure evaluation, the true objective was revealed, and participants were given the option to withdraw with full compensation (600 JPY).
Three participants withdrew, leaving 261 who gave additional consent for the post-disclosure evaluation.

Our experimental design included 6 acts of writing, 3 scenarios per act, and 11 levels of disclosed AI authorship (0\% to 100\% in 10\% increments), resulting in 198 unique conditions. We aimed to collect at least 5 responses per condition, so a total of $6~\text{acts} \times 3~\text{scenarios} \times 11~\text{AI levels} \times 5~\text{response samples} = 990~\text{responses}$.

Each participant evaluated four randomly selected conditions, yielding 1,044 collected responses. To ensure a balanced dataset across all conditions, we randomly down-sampled any condition with more than five responses to exactly five, resulting in a final dataset of 990 responses.

\subsection{Data Analysis}\label{sec:data-analysis}
We used a mixed-methods approach to analyze the data, combining quantitative modeling of rating shifts with qualitative analysis of participants' open-ended explanations.

\subsubsection{Quantitative Analysis}
To analyze how the disclosure of AI authorship affected participants' impressions, we fit a series of linear mixed-effects models. The dependent variables were the five metrics capturing the change in perception, calculated as score differences (i.e., post-disclosure evaluation $-$ pre-disclosure evaluation): \textit{TrustworthinessShift}, \textit{CaringShift}, \textit{CompetenceShift}, \textit{LikabilityShift}, and \textit{FutureShift}.

The independent variables were:
\begin{itemize}
\item \textit{Act}: The six acts of writing--\textit{Convince}, \textit{Interact}, \textit{Reflect}. \textit{Describe}, \textit{Explore}, and \textit{Imagine}--with \textit{Describe} set as the reference category due to its objective and emotionally neutral nature.
\item \textit{AIRatio}: The disclosed percentage of simulated AI authorship in the text (0\%--100\%).
\item \textit{ApplyAI}, \textit{DetectAI}, \textit{PersuasionAI}: Participants' scores on the three MAILS AI literacy sub-scales.
\end{itemize}

Our models included interaction terms between \textit{AIRatio} and all other independent variables. We integrated participant-level random effects to account for individual differences in rating behavior and used the Akaike Information Criterion (AIC) and Likelihood Ratio Test (LRT) for model selection.

\subsubsection{Qualitative Analysis}

We analyzed the 990 open-ended responses using a multi-stage process combining computational topic discovery and manual thematic coding.

First, we pre-processed the data by filtering out 61 non-substantive comments (e.g., ``none,'', ``no change''), leaving 929 responses.
We then tokenized the text using fugashi~\cite{mccann2020fugashi} and applied a custom stopword list to remove low-information words.
We then used this cleaned data for an initial phase of inductive thematic discovery. 
We calculated multilingual Sentence-BERT embeddings~\cite{reimers2020making} and used BERTopic~\cite{grootendorst2022bertopic} for embedding-based clustering with c-TF-IDF to identify potential topic structures. 
To determine the optimal number of topics k, we assessed topic coherence (\(c_v\)) for values ranging from 3 to 30. We selected \(k=14\) which yielded the highest score.

This automated step produced a preliminary thematic structure, but manual review revealed semantic overlaps and misclassified responses.
We therefore proceeded to a second phase of manual refinement.
We merged and redefined the initial 14 topics to create a more robust and unambiguous codebook consisting of 9 final themes (see Table~\ref{tab:topic_model}).
With this finalized codebook, we obtained an initial categorization of all 929 responses using GPT-5. Two authors independently reviewed this classification, resolving disagreements through discussion.

Finally, to characterize each theme, we extracted representative key phrases from the original Japanese responses using 2-grams and 3-grams.
For each theme, we concatenated all associated responses and identified the top n-grams based on their TF-IDF scores.
These n-grams were then reviewed and combined to create meaningful descriptors for each theme.

We conducted the analysis described above using several Python libraries including \texttt{sentence-transformers}, \texttt{BERTopic}, \texttt{scikit-learn} (KMeans), \texttt{fugashi}, \texttt{pandas}, and \texttt{statsmodels}.

\section{Results}

We first present the quantitative findings from our linear mixed-effects models, followed by a thematic analysis of participants' qualitative explanations for their rating changes.

\subsection{Effect of Communicative Purpose on Reader Perception (\ref{rq:act-of-writing})}

We computed marginal and conditional $R^2$ to evaluate the explanatory power of the linear mixed-effect models. 
The marginal $R^2$ was approximately 0.20 across all models, indicating that the fixed effects accounted for a moderate amount of variance. The conditional $R^2$  was approximately 0.30, confirming that the inclusion of participant-level random effects improved the model fit.

\begin{figure*}[p]
    \centering
    \begin{subfigure}[t]{0.47\textwidth}
        \centering
        \includegraphics[width=\textwidth,height=0.28\textheight,keepaspectratio]{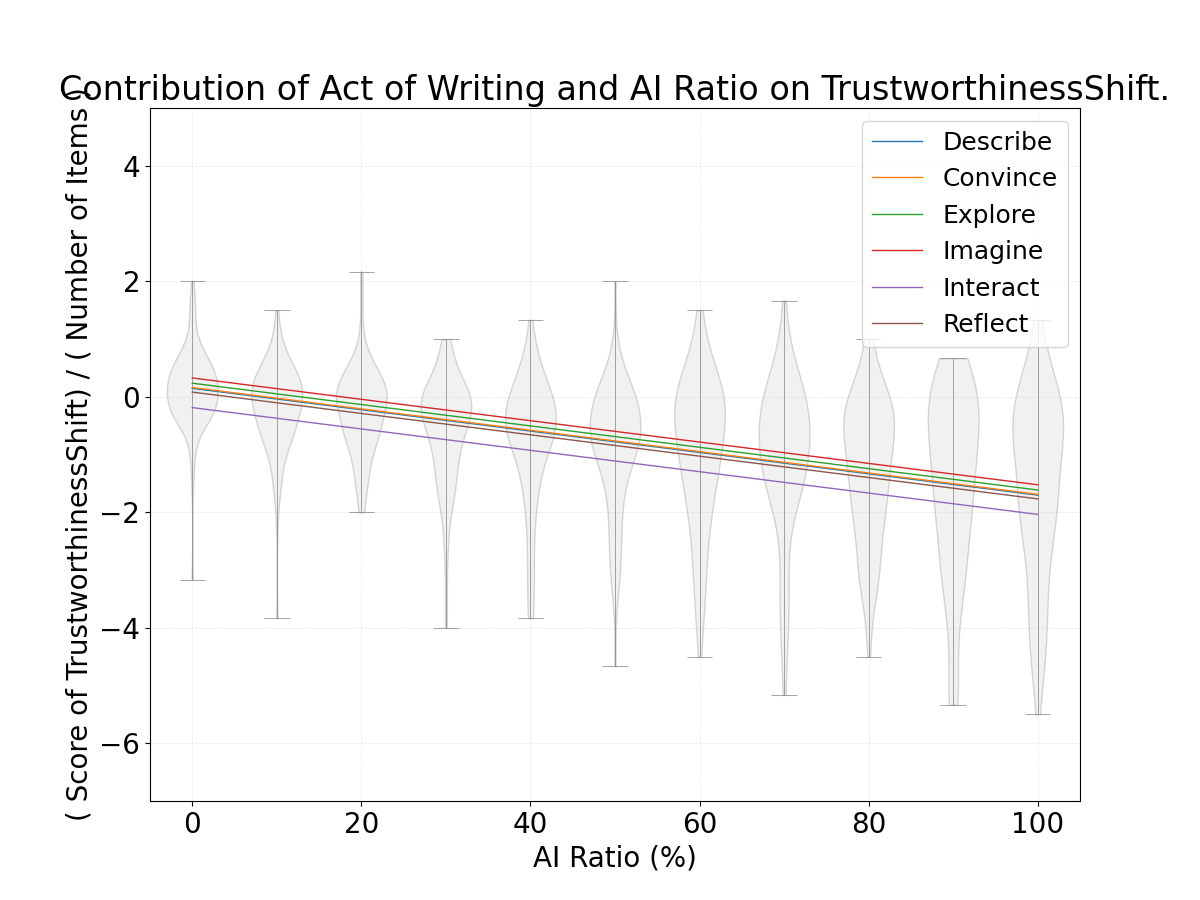}
        \caption{\textit{TrustworthinessShift}}
        \label{fig:regression_visualization_trustworthiness}
    \end{subfigure}
    \hfill
    \begin{subfigure}[t]{0.47\textwidth}
        \centering
        \includegraphics[width=\textwidth,height=0.28\textheight,keepaspectratio]{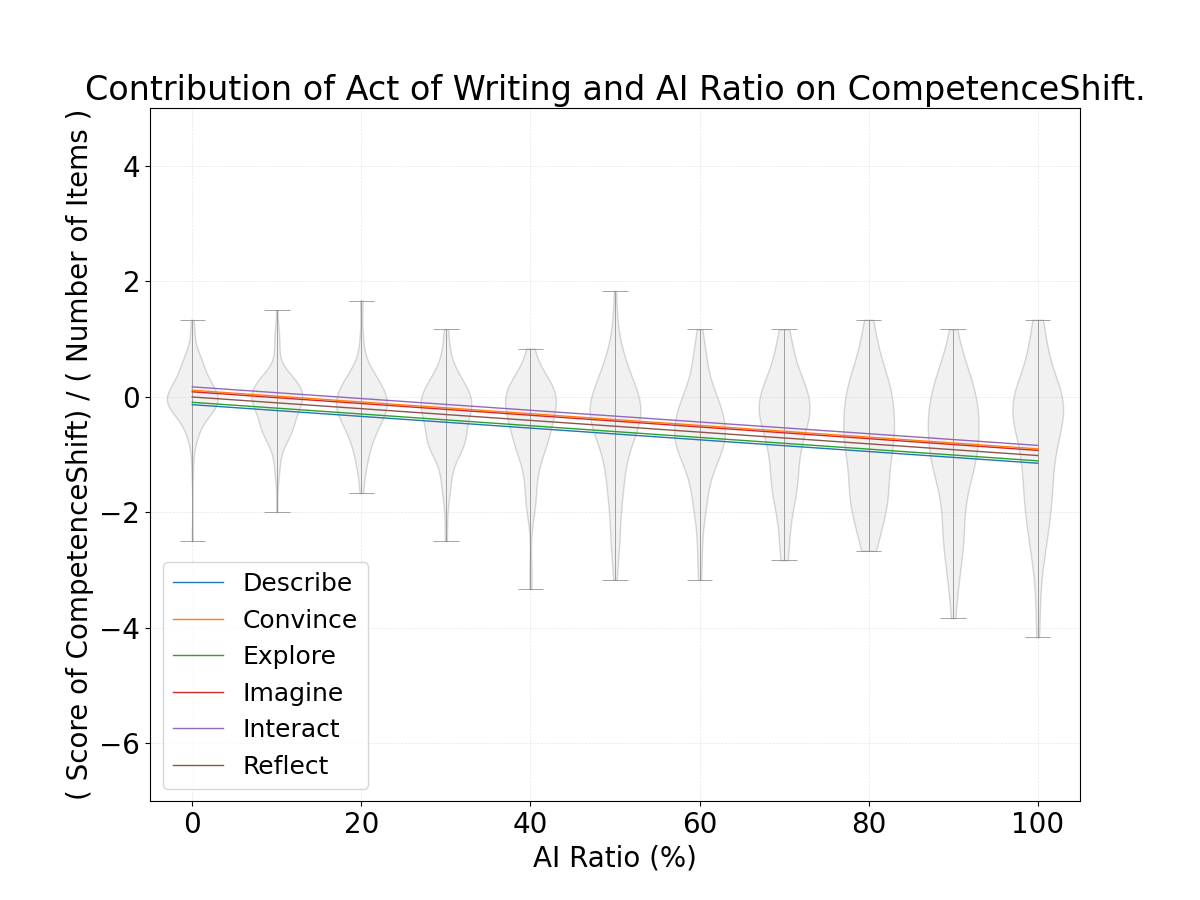}
        \caption{\textit{CompetenceShift}}
        \label{fig:regression_visualization_competence}
    \end{subfigure}

    \begin{subfigure}[t]{0.47\textwidth}
        \centering
        \includegraphics[width=\textwidth,height=0.28\textheight,keepaspectratio]{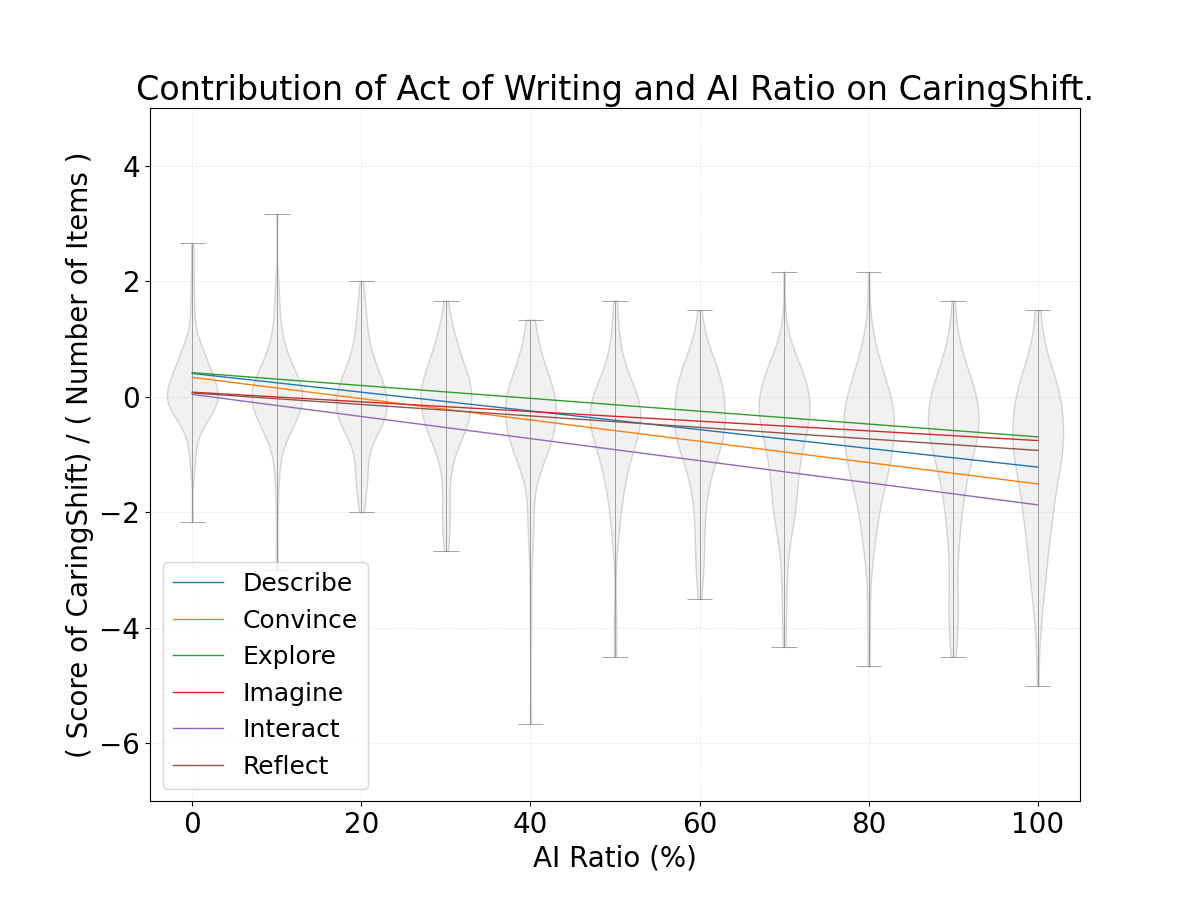}
        \caption{\textit{CaringShift}}
        \label{fig:regression_visualization_caring}
    \end{subfigure}
    \hfill
    \begin{subfigure}[t]{0.47\textwidth}
        \centering
        \includegraphics[width=\textwidth,height=0.28\textheight,keepaspectratio]{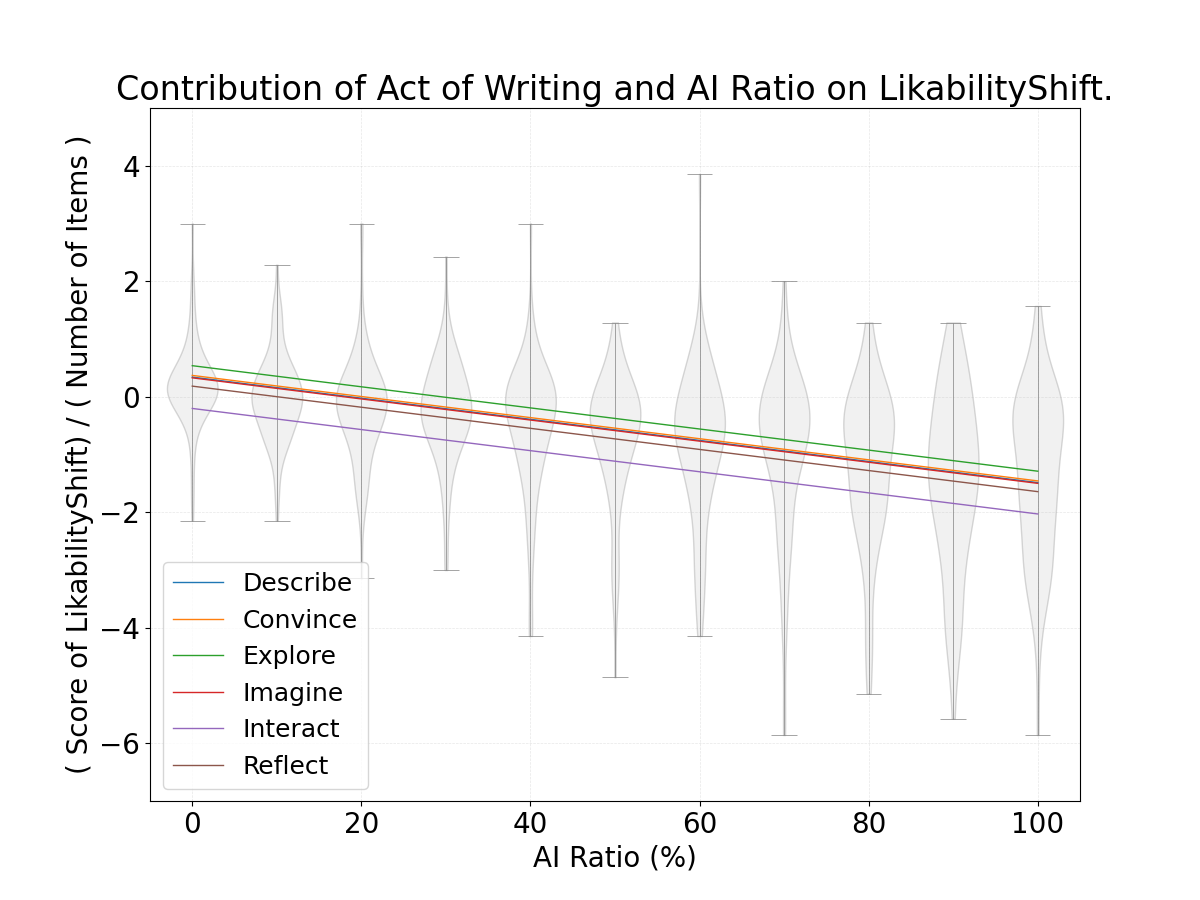}
        \caption{\textit{LikabilityShift}}
        \label{fig:regression_visualization_likability}
    \end{subfigure}

    \begin{subfigure}[t]{0.47\textwidth}
        \centering
        \includegraphics[width=\textwidth,height=0.28\textheight,keepaspectratio]{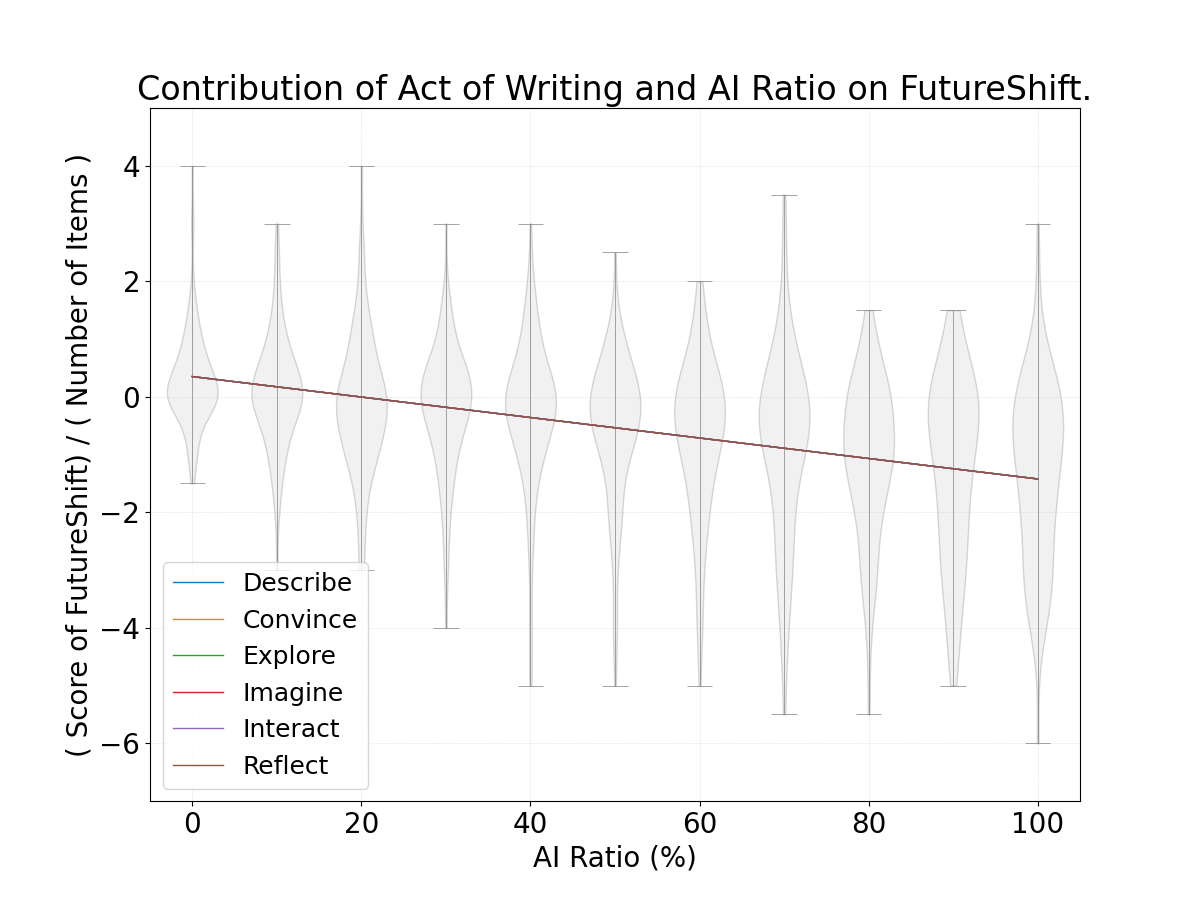}
        \caption{\textit{FutureShift}}
        \label{fig:regression_visualization_future}
    \end{subfigure}

    \caption{
    \textit{Act}-level visualization of the regression analysis for all outcome variables. 
    For visibility, the score of the perception shift was divided by the number of items to uniformize the range across the metrics.
    \textit{AIRatio} shows a negative slope across all metrics. 
    Notable findings: 
    (1) \textit{Interact} has a significantly negative intercept for \textit{TrustworthinessShift}, \textit{CaringShift}, and \textit{Likability}, while it has a significantly positive intercept for \textit{CompetenceShift}.
    (2) \textit{Convince} and \textit{Imagine} have significantly positive intercepts for \textit{CompetenceShift}. 
    (3) \textit{Explore} shows a significantly positive effect for \textit{CaringShift}, and \textit{Imagine} shows a significantly positive conditional effect with \textit{AIRatio} for \textit{CaringShift}.}
    
    \Description{
    \textit{Act}-level visualization of the regression analysis for all outcome variables. 
    For visibility, the score of the perception shift was divided by the number of items to uniformize the range across the metrics.
    \textit{AIRatio} shows a negative slope across all metrics. 
    Notable findings: 
    (1) \textit{Interact} has a significantly negative intercept for \textit{TrustworthinessShift}, \textit{CaringShift}, and \textit{Likability}, while it has a significantly positive intercept for \textit{CompetenceShift}.
    (2) \textit{Convince} and \textit{Imagine} have significantly positive intercepts for \textit{CompetenceShift}. 
    (3) \textit{Explore} shows a significantly positive effect for \textit{CaringShift}, and \textit{Imagine} shows a significantly positive conditional effect with \textit{AIRatio} for \textit{CaringShift}.}
    \label{fig:regression_visualization_all}
\end{figure*}

Table~\ref{tab:summarized_coefficient_matrix} summarizes the key coefficients, with full results provided in the Appendix (Tables~\ref{tab:trustworthiness_mlm}--\ref{tab:future_mlm}). 
Figure~\ref{fig:regression_visualization_all} shows perception shifts by \textit{AIRatio} and \textit{Act}, grouped by reported AI literacy. 
The \textit{Describe} act serves as the reference category for estimating coefficients for the other five communicative acts. All models use \textit{AIRatio} and \textit{Act} as independent variables.

\captionsetup{aboveskip=2pt, belowskip=2pt}

\textit{AIRatio} exhibited a significant negative correlation across all models~($p < .001$). As the disclosed percentage of AI involvement increased, perceptions of the author declined significantly across all five metrics.
The communicative purpose (\textit{Act}) of the text also had a significant influence, with effects varying across the different perception metrics. 
\textit{Interact} had a significant negative effect on \textit{TrustworthinessShift}, \textit{CaringShift}, and \textit{LikabilityShift} ($p < .001$ for all of them).
This suggests AI authorship is negatively viewed for social and interpersonal communication.
In contrast, AI disclosure was viewed more favorably in terms of perceived competence and caring for certain acts.
\textit{Convince}, \textit{Interact}, and \textit{Imagine} were positively associated with \textit{CompetenceShift} ($p < .01$ for \textit{Convince} and \textit{Imagine}, and $p < .001$ for \textit{Interact}).
\textit{Explore} was linked to a more positive shift in perceived caring (\textit{CaringShift}, $p < .01$).
The model also revealed a significant positive interaction between \textit{AIRatio} and \textit{Imagine} for \textit{CaringShift} ($p < .05$), suggesting that the negative effect on perceived caring was mitigated as the stated AI contribution increased in imaginative writing. 

Collectively, these results indicate that readers were more tolerant of AI authorship in argumentative, creative, and exploratory writing compared to interpersonal communication.

\begin{table*}[t]
    \small
    \centering
    \caption{Coefficient matrix summarizing the results of the linear mixed-effects models predicting the shift in author perception based on the disclosed AIRatio, the act of writing, and the reader's AI literacy. Significant effects are highlighted with color, orange for positive and blue for negative effects.}
    \begin{tabular}{lrrrrrr}
    \toprule
    Variables & Trustworthiness & Caring/Goodwill & Competence & Likability & Future \\
    \hline
    \hline
    \multicolumn{6}{l}{\textbf{Main Effect}} \\
    Intercept 
        & \color{teal} \textbf{-4.676***} 
        & \color{teal} \textbf{-2.446***} 
        & \color{teal} \textbf{-3.866***} 
        & \color{teal} \textbf{-4.013***} 
        & \color{teal} \textbf{-1.074***}\\
    AIRatio 
        & \color{teal} \textbf{-3.236***} 
        & \color{teal} \textbf{-2.838***} 
        & \color{teal} \textbf{-1.774***} 
        & \color{teal} \textbf{-3.729***} 
        & \color{teal} \textbf{-1.034***}\\
    \multicolumn{6}{l}{Act (Ref: To Describe)} \\
        \quad To Convince 
            & 0.085 
            & -1.075 
            & \color{redorange} \textbf{1.496**}
            & 0.206 
            & --- \\
        \quad To Explore 
            & 0.536 
            & \color{redorange} \textbf{1.612**} 
            & 0.247 
            & 1.384 
            & --- \\
        \quad To Imagine 
            & 1.080 
            & 0.410 
            & \color{redorange} \textbf{1.335**} 
            & -0.090 
            & --- \\
        \quad To Interact 
            & \color{teal} \textbf{-1.999**} 
            & \color{teal} \textbf{-3.050***} 
            & \color{redorange} \textbf{1.853***} 
            & \color{teal} \textbf{-3.800***} 
            & --- \\
        \quad To Reflect 
            & -0.388 
            & -0.144 
            & 0.801 
            & -1.094 
            & --- \\
    \multicolumn{6}{l}{AI Literacy} \\
        \quad Apply AI Literacy
            & 0.523 
            & --- 
            & 0.220 
            & --- 
            & ---\\
        \quad Detect AI Literacy
            & 0.473 
            & 0.511 
            & 0.361 
            & \color{redorange} \textbf{0.854**} 
            & ---\\
        \quad AI Persuasion Literacy
            & --- 
            & -0.001 
            & --- 
            & --- 
            & 0.105\\
    \addlinespace 
    \multicolumn{6}{l}{\textbf{Interaction}} \\
    \multicolumn{6}{l}{Interaction between AIRatio and Act (Ref: To Describe)} \\
        \quad AIRatio : To Convince
            & --- 
            & -0.390 
            & --- 
            & --- 
            & ---\\
        \quad AIRatio : To Explore
            & --- 
            & 0.892 
            & --- 
            & --- 
            & ---\\
        \quad AIRatio : To Imagine
            & --- 
            & \color{redorange} \textbf{1.372*} 
            & --- 
            & --- 
            & ---\\
        \quad AIRatio : To Interact
            & --- 
            & -0.508 
            & --- 
            & --- 
            & ---\\
        \quad AIRatio : To Reflect
            & --- 
            & 1.098 
            & --- 
            & --- 
            & ---\\
    \multicolumn{6}{l}{Interaction between AIRatio and AI Literacy} \\
        \quad AIRatio : Apply AI Literacy 
            & --- 
            & --- 
            & \color{redorange} \textbf{0.478**} 
            & --- 
            & ---\\
        \quad AIRatio : Detect AI Literacy 
            & \color{redorange} \textbf{0.630**} 
            & --- 
            & --- 
            & \color{redorange} \textbf{0.731**} 
            & ---\\
        \quad AIRatio : AI Persuasion Literacy 
            & --- 
            & \color{redorange} \textbf{0.467*} 
            & --- 
            & --- 
            & \color{redorange} \textbf{0.224**}\\
    \bottomrule
    \multicolumn{6}{l}{\footnotesize * p < 0.05, ** p < 0.01, *** p < 0.001. Dashed lines represent omitted variables through the model selection process.}
    \end{tabular}
    \label{tab:summarized_coefficient_matrix}
\end{table*}

\subsection{Effect of AI Literacy on Reader Perception~(\ref{rq:ai-literacy})}\label{sec:result-quantitative-literacy}

As shown in Table~\ref{tab:summarized_coefficient_matrix}, participants' self-reported AI literacy levels, along with their interactions with \textit{AIRatio}, were significant predictors of perception shifts.
Specifically, higher \textit{ApplyAI} scores were associated with a positive shift in perceived competence (\textit{CompetenceShift}) and had a significant positive interaction with \textit{AIRatio} on this metric ($p < .05$).
\textit{DetectAI} demonstrated significant positive main and interaction effects on the shift in author likability (\textit{LikabilityShift}, $p < .01$ for both), 
as well as a significant positive interaction effect on the shift in trustworthiness (\textit{TrustworthinessShift}, $p < .01$).
Finally, the interaction between \textit{PersuasionAI} and \textit{AIRatio} had a significant positive influence on both the shifts in caring (\textit{CaringShift}, $p < .05$) and desire for future interaction (\textit{FutureShift}, $p < .01$). 
Overall, participants with higher AI literacy were more tolerant, and sometimes even appreciative, of AI use in writing.

\begin{table*}[t]
 \small
 \centering
 \caption{Thematic analysis of open-ended responses explaining changes in reader perception. Abbreviations: Conv.=Convince, Desc.=Describe, Expl.=Explore, Imag.=Imagine, Inter.=Interact, Refl.=Reflect.}
 \begin{tabular}{lp{0.18\linewidth}p{0.22\linewidth}|cccccc|c}
 \toprule
 \textbf{ID} & \textbf{Theme} & \textbf{Representative Words/Phrases} & \multicolumn{6}{c|}{\textbf{Occurrences per Act of Writing}} & \textbf{Total}\\
  &  &  & Conv. & Desc. & Expl. & Imag. & Inter. & Refl. & \\
 \hline
 \hline
 \multicolumn{10}{l}{\textbf{Negative reasons}} \\ \hline 
 N1 & Loss of Human Touch and Sincerity & \textit{Emotion, Feeling, Heart, Compassion, Included, Warmth} & 19 & 7 & 9 & 8 & 22 & 6 & 71 \\
 N2 & Stylistic Awkwardness & \textit{Mechanical, Inanimate, Cold, Unnatural} & 21 & 22 & 22 & 27 & 21 & 19 & 132 \\
 N3 & Diminished Credibility and Expertise & \textit{Credibility, Expertise, Knowledge, Writer, Author, Decrease} & 18 & 39 & 37 & 15 & 7 & 28 & 144 \\
 N4 & Inappropriate Contexts for Using AI & \textit{Apology, Letter, Gratitude, Encouragement, Insincerity} & 25 & 4 & 5 & 8 & 29 & 11 & 82 \\
 N5 & Lack of Human Effort and Agency & \textit{By oneself, Dependence, Get lazy, Writer} & 16 & 11 & 19 & 24 & 10 & 23 & 103 \\
 \hline
 \multicolumn{10}{l}{\textbf{Positive reasons}} \\ \hline 
 P6 & Appropriate and Assistive AI Use & \textit{No awkwardness, No impression change} & 27 & 36 & 33 & 35 & 29 & 23 & 183 \\
 P7 & Improved Quality and Readability & \textit{Utilization, To add, Richness, Intellectual, Easy to read, Better} & 8 & 15 & 14 & 16 & 18 & 16 & 87 \\
 P8 & Positive Attitudes toward AI & \textit{Interesting, Amazing, Surprise, Well done} & 10 & 11 & 11 & 14 & 8 & 15 & 69 \\
 P9 & No AI Authorship & \textit{Favorable, Trust, Honesty, Not using} & 7 & 10 & 8 & 10 & 11 & 12 & 58 \\
 \bottomrule
 \end{tabular}
 \label{tab:topic_model}
\end{table*}

\subsection{Qualitative Reasons for Perception Shifts (\ref{rq:perception})}

In our thematic analysis of the 929 open-ended responses, we identified nine themes: five explaining negative shifts and four explaining positive or neutral ones. These themes are summarized in Table~\ref{tab:topic_model}.

\subsubsection{Negative Themes: Why Perceptions Worsened}

\paragraph{\underline{Theme N1: Loss of Human Touch and Sincerity} (71 responses)}

This theme captures a perceived loss of authenticity and emotional connection. Participants reported that disclosure drained the text of its warmth, even when they had initially found it compelling. One commented: \textit{Even though the writing is skillful, it doesn't feel heartfelt''} (\textit{Act}: Imagine, \textit{AIRatio}: 30\%). Another explained: \textit{The text seemed well-written, but learning it was from an AI made it feel like it lacked a soul, so I lowered my evaluation''} (\textit{Act}: Convince, \textit{AIRatio}: 20\%). These comments highlight that for many readers, AI authorship severs the link to a human author's perceived emotions and sincerity, undermining the relational value of the text.

\paragraph{\underline{Theme N2: Stylistic Awkwardness} (132 responses)}

This theme centers on linguistic cues that participants associated with AI, such as a ``mechanical'' or ``unnatural'' style. As one participant noted, \textit{It felt more mechanical, which changed my impression of its warmth and friendliness''} (\textit{Act}: Convince, \textit{AIRatio}: 100\%). Another pointed out the importance of human oversight: \textit{I felt there were too many unnatural expressions. Using AI can be good, but without proper quality control, it can have the opposite effect''} (\textit{Act}: Reflect, \textit{AIRatio}: 40\%). This theme shows that beyond a general feeling of inauthenticity, specific stylistic artifacts perceived as non-human can directly trigger negative judgments.

\paragraph{\underline{Theme N3: Diminished Author Credibility and Expertise} (144 responses)}

This was the most prevalent negative theme. Participants questioned the author's credibility and expertise upon learning of AI's role, expressing distrust in the content. As one participant noted: \textit{It seemed like AI explained most of the reasoning, so I found it hard to trust the author. It's better to limit AI use to just making the writing clearer''} (\textit{Act}: Explore, \textit{AIRatio}: 30\%).
In some cases, the perceived competence of the AI overshadowed that of the human: \textit{``I got the sense that AI handled some of the more technical parts... I was surprised by how intelligent the AI was, enough to actually change my impression of the writer''} (\textit{Act}: Describe, \textit{AIRatio}: 50\%).
This theme reveals a key tension: while participants see value in using AI for information gathering, they grow skeptical of the author’s diligence and expertise when the role of AI appears too substantial, particularly if they suspect a lack of fact-checking. One participant remarked: \textit{``Using AI to obtain specialized information is very effective, but since it’s unclear whether the author has actually verified that information, it becomes very difficult to trust their judgment''} (\textit{Act}: Describe, \textit{AIRatio}: 80\%).
This suggests that when AI is perceived to contribute core knowledge, it can undermine the author's perceived competence.

\paragraph{\underline{Theme N4: Inappropriate Contexts for Using AI} (82 responses)}

Participants felt that AI was simply inappropriate for certain communicative acts, particularly those with strong social and emotional stakes. This sentiment was most common for acts like \textit{Convince} (25 occurrences) and \textit{Interact} (29 occurrences). For example, regarding an apology letter, a participant wrote: \textit{"If someone uses AI, even just a little, to write an apology letter, I can’t help but question whether it’s truly sincere. For that reason, I rated their trustworthiness and sincerity lower"} (\textit{Act}: Convince, \textit{AIRatio}: 10\%). Similarly, for a letter of encouragement, another stated, \textit{"I didn’t need a long letter written with AI — I just wanted some words of encouragement in their own voice"} (\textit{Act}: Interact, \textit{AIRatio}: 30\%). 
This theme strongly suggests that in emotionally and relationally significant contexts, AI authorship is often perceived as a violation of social expectations.

\paragraph{\underline{Theme N5: Lack of Human Effort and Agency} (103 responses)}
Finally, many participants viewed extensive AI use as a sign of laziness or a lack of personal effort from the author. Participants felt the author was outsourcing their responsibility to think and write. One commented, \textit{"Using AI to write gives the impression of just being lazy"} (\textit{Act}: Convince, \textit{AIRatio}: 40\%).
Others felt it diminished the author's voice: \textit{``Since AI was used for about 30\% of the writing, I felt the author’s own intention came through a bit less''} (\textit{Act}: Reflect, \textit{AIRatio}: 30\%). 
This judgment became more severe as the disclosed proportion of AI authorship increased. A participant who saw a 70\% disclosure commented: \textit{``They’re relying way too much on AI, and there’s barely any effort to proofread what the AI wrote. I can’t help but lower my evaluation of them as a person''} (\textit{Act}: Describe, \textit{AIRatio}: 70\%). Notably, 77 of the 103 responses in this theme corresponded to conditions where the disclosed \textit{AIRatio} was 50\% or higher, showing that readers penalize what they perceive as an over-reliance on automation.

\subsubsection{Positive and Neutral Themes}

\paragraph{\underline{Theme P6: Appropriate and Assistive AI Use} (183 responses)}

The most frequent positive theme was the perception that AI was used in an appropriate, assistive, or supplementary capacity. This was particularly true when the disclosed AI contribution was low. 125 of these responses (68\%) were for texts with an \textit{AIRatio} under 50\%. In such cases, participants felt the author remained in control, as one noted: \textit{``Since only about 10\% of the text was edited by AI, and it was an explanation of how to operate industrial equipment, it felt like the AI simply identified and filled in missing parts of what a human had written. So, knowing that AI was used didn’t change my evaluation''} (\textit{Act}: Describe, \textit{AIRatio}: 10\%). AI was seen as a helpful partner rather than a replacement: \textit{``AI supported the writing nicely, which slightly improved my evaluation. It gave a friendly impression, as if the AI were helping the human''} (\textit{Act}: Reflect, \textit{AIRatio}: 30\%). Others saw a constructive partnership: \textit{``In the parts handled by AI, I got the impression that there were more technical terms. I felt like AI took charge of the technical side while the human handled the emotional side''} (\textit{Act}: Reflect, \textit{AIRatio}: 40\%). This suggests that when AI is perceived as a tool that supports rather than supplants the author, its use is deemed acceptable and even beneficial.

\paragraph{\underline{Theme P7: Improved Quality and Readability} (87 responses)}

This theme reflects an appreciation for the ability of AI to improve the quality and clarity of the writing. Participants valued the contribution of AI when it led to clearer explanations and better structure.
One commented: \textit{``It felt like the AI provided clear and natural supplementary explanations that made the text easier to understand''} (\textit{Act}: Describe, \textit{AIRatio}: 50\%). Another appreciated the impact on conciseness: \textit{``It seems that the sentences became shorter after the AI’s editing... Having shorter sentences makes it easier to follow''} (\textit{Act}: Explore, \textit{AIRatio}: 70\%).
This appreciation extended to complex and creative tasks, where AI was seen as a valuable tool for articulating ideas more effectively: \textit{``Organizing and expressing complex subjects objectively isn’t easy, [so] if AI can help with that, it’s actually valuable''} (\textit{Act}: Explore, \textit{AIRatio}: 60\%).
For a creative piece, one participant remarked, \textit{``Without the parts added by AI, the story would have felt shallow and lacking substance. With the AI’s support, it turned into a much better story''} (\textit{Act}: Imagine, \textit{AIRatio}: 60\%). This theme shows that AI's contribution was valued when it was perceived to enhance the author's message without replacing their core intent.

\paragraph{\underline{Theme P8: Positive Attitudes toward AI} (69 responses)}
This theme was characterized by participants' general surprise, interest, or admiration for AI's capabilities. 
Comments were often brief and expressive of a ``wow'' factor: \textit{``It’s unbelievably well done''} (\textit{Act}: Reflect, \textit{AIRatio}: 90\%) and \textit{``My overall evaluation didn’t really change, but I was impressed by how knowledgeable the AI seemed''} (\textit{Act}: Describe, \textit{AIRatio}: 30\%). 
For some, this positive reaction translated into a desire to engage with AI themselves: \textit{``This is the first time I read a poem written by AI. Next time, I’d like to try writing a poem with the help of AI''} (\textit{Act}: Imagine, \textit{AIRatio}: 70\%). 
Participants in this category tended to have higher self-reported AI literacy. 
Their average scores for \textit{ApplyAI}, \textit{DetectAI}, and \textit{PersuasionAI} were 9.9 ($SD$=5.5), 2.2 ($SD$=4.2), and 2.7 ($SD$=3.9), respectively, compared to the overall sample averages of 4.6 ($SD$=7.7), -1.5 ($SD$=3.9), and -0.5 ($SD$=3.4).
This aligns with our quantitative finding that higher AI literacy correlates with more positive reactions to AI authorship. 
Interestingly, the frequent expressions of surprise from this more AI-literate subgroup suggest that the capabilities of current generative models continue to exceed the expectations of even relatively experienced users.

\paragraph{\underline{Theme P9: Positive Reaction to No AI Authorship} (58 responses)}

This theme exclusively comprises reactions to the 0\% \textit{AIRatio} condition, where participants were told the text was entirely human-written. While some simply stated that their evaluation did not change, many expressed increased appreciation for the author's skill and sincerity upon this confirmation. One participant wrote, \textit{``Since the report was written entirely by a human without using AI, I felt (the author was) very intelligent, capable, and sincere, so I rated it highly''} (\textit{Act}: Explore, \textit{AIRatio}: 0\%).  This suggests that explicitly labeling a text as "human-written" can enhance an author's standing, akin to the premium placed on products marketed as ``handmade''.
\section{Discussion}

Our findings provide a nuanced understanding of how readers' perceptions of an author shift upon the disclosure of AI authorship. We distill our results through the lens of our three research questions, discussing the significance of how communicative purpose, AI literacy, and perceived authorial effort shape reader judgments.

\subsection{RQ1: Context is Critical: AI for Creation, Not Connection}
Our results reveal a sharp distinction in how readers perceive AI authorship depending on the communicative purpose of the text (Table~\ref{tab:summarized_coefficient_matrix}). In acts of writing that are object-oriented, such as persuading (\textit{Convince}), creating narratives (\textit{Imagine}), or developing knowledge (\textit{Explore}), AI involvement was viewed more favorably. 
This task-dependent favoritism aligns with insights from \citet{castelo2019task}, who found that objective tasks, such as predicting stock market outcomes, were more readily accepted than subjective tasks like romantic partner recommendation. However, AI generation may still be discouraged in objective tasks where accuracy is paramount, particularly in forecasting contexts~\cite{dietvorst2015algorithm}.
The positive association with perceived competence in \textit{Convince} and \textit{Imagine} suggests that readers may appreciate AI as a tool that enhances the quality of argumentation or creative expression. 
This is supported by prior work suggesting that readers value AI's contribution when it serves to consolidate or develop an author's ideas~\cite{hwang202580}.

In stark contrast, the use of AI in the person-oriented act of \textit{Interact} elicited strong negative reactions. Our qualitative themes confirm that in such contexts, AI-authored text was perceived as ``cold'', ``impersonal'' (Theme N1), and even ``insincere'' (Theme N4). 
This negative perception is consistent with prior work showing that AI disclosure can signal a lack of author engagement and thereby erode credibility~\cite{jain2024revealing, jakesch2019ai, schilke2025transparency}. 
These findings underscore the context sensitivity of algorithm aversion.
As noted by \citet{morewedge2022preference}, algorithm aversion is often triggered by tasks lacking established evaluation criteria (e.g., creative work) or those dependent on individual subjectivity (e.g., gift recommendations).
The \textit{Interact} act embodies both characteristics, and our findings confirm it to be the act of writing most sensitive to AI disclosure.
This also highlights the importance of the ``communicative relationship'' in interpersonal communication~\cite{burleson2010nature}. Because AI is perceived as incapable of genuine empathetic engagement~\cite{liu2022will, rubin2025comparing}, its use in relational contexts can be interpreted as a violation of social expectations, making human attribution critical. 
Interestingly, in the \textit{Convince} and \textit{Interact} act, AI use boosted perceived competence, suggesting that when AI's role is to strengthen an argument and leverage its perceived objectivity~\cite{ovsyannikova2025third}, its use can be seen as strategic rather than socially detrimental.

\subsection{RQ2: Moderating Roles of AI Authorship Level and Readers' Literacy}
Two factors emerged as key determinants of the magnitude of perception shifts: the disclosed level of AI contribution and the reader's AI literacy. As confirmed by our regression analysis, a higher disclosed \textit{AIRatio} consistently led to more negative perceptions of the author. Our qualitative themes provide the reasoning behind this trend: as AI's role grew, participants were more likely to perceive a loss of human touch (Theme N1), question the author's expertise (Theme N3), and infer a lack of effort or agency (Theme N5). This echoes recent work suggesting that dominant AI authorship can lower an author's perceived credibility and ownership~\cite{draxler2024ai, kirk2025ai}.

Our results also confirm that a reader's AI literacy can significantly mitigate these negative effects. Participants with higher self-reported AI literacy exhibited smaller negative perception shifts and, in some cases, expressed positive attitudes toward AI's capabilities (Theme P8). 
This is consistent with prior studies indicating that greater familiarity with AI and positive past interactions with AI foster a more pragmatic and less critical evaluation of its use~\cite{bohm2023people, li2024does, mahmud2024decoding}. 
AI literacy appears to function as a calibration mechanism, allowing readers to more accurately assess the role and utility of AI in the writing process. However, the effect size of AI literacy was smaller than that of the writing act itself. This implies that while literacy can soften negative reactions, it does not erase the strong social norms governing contexts like interpersonal communication, where even AI-literate readers often expect human-authored content.

\subsection{RQ3: Assistive AI Authorship Can Be Socially Acceptable}
While the overall trend in our study was toward negative perception shifts, a substantial portion of responses (38.6\%) indicated neutral or even positive changes. These reactions were most common when the disclosed AI contribution was low (under 50\%) and occurred in descriptive or argumentative acts such as \textit{Explore}, \textit{Convince}, and \textit{Imagine}, rather than interpersonal acts like \textit{Interact}.

Our qualitative data revealed that such assistive AI authorship was perceived positively when it served a clear, assistive role that did not supplant the author's voice. 
This preference may stem from human favoritism, where human authorship serves as a cue for trust~\cite{buchanan2024people, zhang2023human}.
Participants agreed that when AI was used for supplementary tasks, the author's core intention and argument can be preserved, or even strengthened (Theme P6). 
This aligns with prior findings that limiting a machine's capability to an advisory role can lead to greater permissibility in decision-making processes~\cite{bigman2018people}.
In the context of non-emotional writing, \citet{hwang202580} similarly found that writers perceive their work as authentic when AI is used to refine or develop their own pre-existing ideas.
Our findings confirm that the social acceptability of AI authorship hinges on its perceived role: when positioned as a supportive tool that enhances human expression, it can be valued; when it is seen as a substitute for human effort and emotion, it is often rejected.

\subsection{Design Implications for AI-mediated Writing}

Based on our findings, we formulate four design implications for future AI-mediated writing systems aiming to balance utility with the preservation of authenticity and trust.

\subsubsection{Design for Functional, Context-Sensitive Transparency}
Our results showed that simple quantitative disclosure, such as stating the amount of AI contribution, was a blunt and insufficient instrument, particularly in relational contexts. 
Moreover, recent research suggests that readers often anticipate others will use AI tools more than they themselves~\cite {purcell2024people}.
This social assumption may raise suspicion regarding AI involvement, potentially eroding trust in the author~\cite{jakesch2019ai}.
Consequently, we believe that transparency mechanisms should disclose the functional use of AI in a context-sensitive manner. For example, rather than focusing on \textit{how much} AI was used, systems should help authors explain the \textit{nature} and \textit{purpose} of AI involvement (e.g., ``AI assisted with grammatical revisions'' or ``AI generated the initial draft'').
This aligns with emerging best practices, such as the ACM's policy on authorship, which calls for specifying the role AI played~\cite{ACM2025AuthorshipPolicy}.

This functional transparency is particularly relevant for non-native speakers, who often use AI to ensure linguistic accuracy rather than to offload cognitive effort~\cite{hwang2023chatgpt, li2024exploring}. In such cases, explicitly disclosing that AI was used to overcome language barriers can frame the assistance as a pragmatic accessibility aid. This context may induce reader sympathy and understanding, thereby mitigating the negative impact of disclosure by highlighting the author's effort to communicate effectively in a non-native language.

Furthermore, the mode of disclosure should adapt to the communicative act. 
For descriptive or exploratory writing, explicitly showing the history of AI edits, such as with text highlighting~\cite{hoque2024hallmark}, may help increase author credibility. 
For creative co-writing, disclosure could emphasize the collaborative nature of the process. In sensitive interpersonal contexts, transparency should focus on preserving the human author's effort and emotional investment.

\subsubsection{Preserve and Communicate Human Effort}
Our qualitative analysis revealed that negative reactions to AI disclosure were significantly mitigated when authentic human effort remained perceptible. This suggests that readers evaluate not just the extent of AI authorship, but the degree of perceptible human agency in the final text. 
Therefore, AI writing tools should be designed not only to disclose the role of AI but also to actively preserve and communicate the author's contributions. Interfaces could integrate features that visualize human effort, such as edit histories, authorship heatmaps, or representations of the author's strategic decisions and intentions~\cite{peng2025navigating, wang2025intentprism}. 
By foregrounding the author’s agency (e.g., the acts of guiding, initiating, and curating the text as conceptualized by \citet{rae2024effects}), systems can frame human-AI collaboration as a process of thoughtful engagement rather than delegation. 
This approach complements context-sensitive transparency. As social and emotional contexts place a premium on human effort, systems should explicitly and visibly convey human touch to proactively restore authenticity and trust.

\subsubsection{Provide Adaptive Support in Emotional and Social Contexts}
The strongest negative perception shifts occurred in writing with emotional or relational significance, where readers expect sincerity and personal investment. This underscores that the social function of writing fundamentally shapes how AI authorship is judged. AI-assisted writing systems may therefore benefit from being designed with social and emotional awareness. 
For example, they could detect communicative intent by recognizing linguistic markers of empathy, apology, or gratitude and alert users to the social risks of automation and suggest drafting the content personally. 
Alternatively, it could adapt its support by limiting suggestions to structural or stylistic refinements while leaving core affective expressions to the human author. Systems could also prompt writers to add personal reflections or review the text to ensure it conveys genuine feelings. 
Such adaptive designs would not only help preserve relational authenticity but also guide authors toward a more balanced and appropriate division of labor with AI.

\subsubsection{Foster AI Literacy for Both Authors and Readers}
Our study confirms that higher AI literacy correlates with greater tolerance for AI authorship. This suggests that perceptions are shaped not only by the act of writing but also by the user's understanding of the role of AI. We therefore view enhancing AI literacy as an important step toward fostering calibrated trust. AI writing tools can be designed to actively cultivate this literacy. For authors, reflective interfaces could visualize their reliance patterns, such as the frequency with which they accept suggestions verbatim versus editing them, and prompt them to consider their authorship balance. 
For readers, provenance interfaces could provide contextual information about typical forms of AI assistance, 
helping them calibrate trust without overreacting to the mere fact of AI disclosure, thereby allowing them to better perceive the author's dedication.
Over time, such feedback loops could help establish shared social norms around what constitutes appropriate and considerate use of AI in writing.

\subsection{Limitations and Future Work}
Our study has several limitations that suggest avenues for future research.

First, we manipulated the perceived level of AI authorship using a quantitative ratio of sentences. While systematic, this does not account for the semantic weight of those sentences. For instance, a sentence articulating a core thesis carries more weight than one providing a minor detail; consequently, the perceived influence of AI likely varied depending on which specific sentences were highlighted. Future work should explore how reader perceptions are affected when AI contributes to different functional parts of a text, such as core arguments versus stylistic refinements.

Second, our simulation of AI disclosure simplifies the complexity of real-world human-AI writing collaborations. As suggested by prior work, a more nuanced representation would involve describing the nature of the collaboration, such as the types of prompts used, the specific assistance provided by the AI, and how the author accepted, rejected, or modified those suggestions~\cite{draxler2024ai, he2025contributions, rae2024effects}. Future experiments should include these contribution-level descriptions to better model diverse patterns of human-AI co-authorship.

Third, our study was conducted exclusively with Japanese participants. 
While our design was not language-specific, cultural norms surrounding communication, authenticity, and technology may influence how AI authorship is perceived. 
The generalizability of our findings to other linguistic and cultural contexts remains an open question, and cross-cultural comparative studies are needed to investigate how these nuanced social dynamics vary across different populations.

Finally, we assessed AI literacy as a static personal attribute. However, AI literacy and attitudes toward AI are likely to evolve over time. Future longitudinal research is necessary to determine whether our findings persist as AI technologies and user familiarity continue to mature.

\section{Conclusion}

We examined how the disclosure of AI authorship in writing influences readers’ perceptions across different acts of writing.
Our study with 261 participants revealed that disclosure generally erodes perceptions of trustworthiness, caring, competence, and likability, particularly within social and emotional contexts. However, these negative effects are moderated by the reader's AI literacy and the degree to which human effort and agency remain visible. Our findings demonstrate that the social acceptability of AI-mediated writing hinges not merely on textual quality, but on how human intentionality is communicated and perceived. Based on these insights, we propose design implications for AI writing tools that prioritize context-sensitive transparency, the preservation of human effort, and the cultivation of AI literacy to sustain authenticity. Our work advocates for re-imagining AI writing assistance as a co-creative partnership and relational process where technology amplifies, rather than replaces, human expression and empathy. As AI becomes increasingly embedded in our communicative practices, future research must continue to explore these dynamics across diverse cultural contexts and longitudinal timelines.
\section{GenAI Usage Disclosure}
Generative AI tools were used in several stages of this research, including the creation of experimental materials, data analysis, and manuscript preparation. In all cases, AI-generated output was carefully reviewed, validated, and revised by the authors.

The 18 text stimuli used in the experiment were initially drafted in English using GPT-4o. As detailed in the Method section, these texts were then translated and subsequently reviewed and edited by a native Japanese-speaking author to ensure linguistic naturalness and quality.

For our data analysis, we used AI-powered coding assistants like Cursor to aid in scripting the statistical analyses. In our qualitative analysis, GPT-5 was used to perform an initial categorization of the open-ended responses, which served as a foundation for the subsequent manual thematic analysis. We also used GPT-5 to generate initial English translations of participant quotes from the original Japanese. These translations were then manually verified and refined by an author to ensure fidelity to the original meaning.

Finally, throughout the writing process of this manuscript, we used Gemini, GPT-5, and Writefull to improve the clarity, style, and grammatical correctness of the manuscript.

\begin{acks}

This work was partly supported by JST ASPIRE for Top Scientists (Grant Number JPMJAP2405) and JST PRESTO (Grant Number JPMJPR23IB).

\end{acks}
\bibliographystyle{ACM-Reference-Format}
\bibliography{references}

@STRING{nov = "Nov."}

@STRING{health = "ACM Transactions on Computing for Healthcare"}

@article{hancock2020ai,
  title={AI-mediated communication: Definition, research agenda, and ethical considerations},
  author={Hancock, Jeffrey T and Naaman, Mor and Levy, Karen},
  journal={Journal of Computer-Mediated Communication},
  volume={25},
  number={1},
  pages={89--100},
  year={2020},
  publisher={Oxford University Press}
}

@article{Barkallah2025Transparent,
title={Transparent Hearts: Balancing Privacy and Trust in AI-Generated Self-Presentation for Online Dating},
author={Meryem Barkallah and Yosr Aissa and Douglas Zytko},
journal={Proceedings of the Extended Abstracts of the CHI Conference on Human Factors in Computing Systems},
year={2025},
doi={10.1145/3706599.3720311}
}

@article{mieczkowski2021ai,
  title={AI-mediated communication: Language use and interpersonal effects in a referential communication task},
  author={Mieczkowski, Hannah and Hancock, Jeffrey T and Naaman, Mor and Jung, Malte and Hohenstein, Jess},
  journal={Proceedings of the ACM on Human-Computer Interaction},
  volume={5},
  number={CSCW1},
  pages={1--14},
  year={2021},
  publisher={ACM New York, NY, USA}
}

@article{Miura2025Understanding,
title={Understanding and Supporting Formal Email Exchange by Answering AI-Generated Questions},
author={Y. Miura and Chi-Lan Yang and Masaki Kuribayashi and Keigo Matsumoto and Hideaki Kuzuoka and Shigeo Morishima},
journal={Proceedings of the 2025 CHI Conference on Human Factors in Computing Systems},
year={2025},
doi={10.1145/3706598.3714016}
}

@inproceedings{mirowski2023co,
  title={Co-writing screenplays and theatre scripts with language models: Evaluation by industry professionals},
  author={Mirowski, Piotr and Mathewson, Kory W and Pittman, Jaylen and Evans, Richard},
  booktitle={Proceedings of the 2023 CHI conference on human factors in computing systems},
  pages={1--34},
  year={2023}
}

@article{carolus2023mails,
  title     = {MAILS-Meta AI literacy scale: Development and testing of an AI literacy questionnaire based on well-founded competency models and psychological change-and meta-competencies},
  author    = {Carolus, Astrid and Koch, Martin J and Straka, Samantha and Latoschik, Marc Erich and Wienrich, Carolin},
  journal   = {Computers in Human Behavior: Artificial Humans},
  volume    = {1},
  number    = {2},
  pages     = {100014},
  year      = {2023},
  publisher = {Elsevier}
}

@article{ng2022using,
  title     = {Using digital story writing as a pedagogy to develop AI literacy among primary students},
  author    = {Ng, Davy Tsz Kit and Luo, Wanying and Chan, Helen Man Yi and Chu, Samuel Kai Wah},
  journal   = {Computers and Education: Artificial Intelligence},
  volume    = {3},
  pages     = {100054},
  year      = {2022},
  publisher = {Elsevier}
}

@inproceedings{long2020ai,
  title     = {What is AI literacy? Competencies and design considerations},
  author    = {Long, Duri and Magerko, Brian},
  booktitle = {Proceedings of the 2020 CHI conference on human factors in computing systems},
  pages     = {1--16},
  year      = {2020}
}

@article{wang2023measuring,
  title     = {Measuring user competence in using artificial intelligence: validity and reliability of artificial intelligence literacy scale},
  author    = {Wang, Bingcheng and Rau, Pei-Luen Patrick and Yuan, Tianyi},
  journal   = {Behaviour \& information technology},
  volume    = {42},
  number    = {9},
  pages     = {1324--1337},
  year      = {2023},
  publisher = {Taylor \& Francis}
}

@inproceedings{liu2022will,
  title     = {Will AI console me when I lose my pet? Understanding perceptions of AI-mediated email writing},
  author    = {Liu, Yihe and Mittal, Anushk and Yang, Diyi and Bruckman, Amy},
  booktitle = {Proceedings of the 2022 CHI conference on human factors in computing systems},
  pages     = {1--13},
  year      = {2022}
}

@article{berge2016wheel,
  author    = {Kjell Lars Berge and Lars Sigfred Evensen and Ragnar Thygesen},
  title     = {The Wheel of Writing: a model of the writing domain for the teaching and assessing of writing as a key competency},
  journal   = {The Curriculum Journal},
  volume    = {27},
  number    = {2},
  pages     = {172--189},
  year      = {2016},
  publisher = {Routledge},
  doi       = {10.1080/09585176.2015.1129980},
  url       = {https://doi.org/10.1080/09585176.2015.1129980},
  eprint    = {https://doi.org/10.1080/09585176.2015.1129980}
}

@article{mccroskey1999goodwill,
  author    = {James C. McCroskey and Jason J. Teven},
  title     = {Goodwill: A reexamination of the construct and its measurement},
  journal   = {Communication Monographs},
  volume    = {66},
  number    = {1},
  pages     = {90--103},
  year      = {1999},
  publisher = {NCA Website},
  doi       = {10.1080/03637759909376464},
  url       = {https://doi.org/10.1080/03637759909376464},
  eprint    = {https://doi.org/10.1080/03637759909376464}
}

@article{reysen2005construction,
  title     = {Construction of a new scale: The Reysen likability scale},
  author    = {Reysen, Stephen},
  journal   = {Social Behavior and Personality: an international journal},
  volume    = {33},
  number    = {2},
  pages     = {201--208},
  year      = {2005},
  publisher = {Scientific Journal Publishers}
}

@article{sprecher2021closeness,
  title     = {Closeness and other affiliative outcomes generated from the Fast Friends procedure: A comparison with a small-talk task and unstructured self-disclosure and the moderating role of mode of communication},
  author    = {Sprecher, Susan},
  journal   = {Journal of Social and Personal Relationships},
  volume    = {38},
  number    = {5},
  pages     = {1452--1471},
  year      = {2021},
  publisher = {Sage Publications Sage UK: London, England}
}

@article{grootendorst2022bertopic,
  title   = {BERTopic: Neural topic modeling with a class-based TF-IDF procedure},
  author  = {Grootendorst, Maarten},
  journal = {arXiv preprint arXiv:2203.05794},
  year    = {2022}
}

@article{schilke2025transparency,
  title     = {The transparency dilemma: How AI disclosure erodes trust},
  author    = {Schilke, Oliver and Reimann, Martin},
  journal   = {Organizational Behavior and Human Decision Processes},
  volume    = {188},
  pages     = {104405},
  year      = {2025},
  publisher = {Elsevier}
}

@article{sah2018conflict,
  title     = {Conflict of interest disclosure as an expertise cue: Differential effects due to automatic versus deliberative processing},
  author    = {Sah, Sunita and Malaviya, Prashant and Thompson, Debora},
  journal   = {Organizational Behavior and Human Decision Processes},
  volume    = {147},
  pages     = {127--146},
  year      = {2018},
  publisher = {Elsevier}
}

@inproceedings{jakesch2019ai,
  title     = {AI-mediated communication: How the perception that profile text was written by AI affects trustworthiness},
  author    = {Jakesch, Maurice and French, Megan and Ma, Xiao and Hancock, Jeffrey T and Naaman, Mor},
  booktitle = {Proceedings of the 2019 CHI conference on human factors in computing systems},
  pages     = {1--13},
  year      = {2019}
}

@inproceedings{dugan2023real,
  title     = {Real or fake text?: Investigating human ability to detect boundaries between human-written and machine-generated text},
  author    = {Dugan, Liam and Ippolito, Daphne and Kirubarajan, Arun and Shi, Sherry and Callison-Burch, Chris},
  booktitle = {Proceedings of the AAAI Conference on Artificial Intelligence},
  volume    = {37},
  number    = {11},
  pages     = {12763--12771},
  year      = {2023}
}

@article{hakam2024human,
  title     = {Human-written vs AI-generated texts in orthopedic academic literature: comparative qualitative analysis},
  author    = {Hakam, Hassan Tarek and Prill, Robert and Korte, Lisa and Lovrekovi{\'c}, Bruno and Ostoji{\'c}, Marko and Ramadanov, Nikolai and Muehlensiepen, Felix},
  journal   = {JMIR formative research},
  volume    = {8},
  pages     = {e52164},
  year      = {2024},
  publisher = {JMIR Publications Toronto, Canada}
}

@article{wang2025chatgpt,
  title     = {When ChatGPT Speaks About Health: Examining Perceptions of Warmth and Competence Toward AI as a Health Information Source},
  author    = {Wang, Buduo and Shibo, Bruce Wang and Kafle, Jiwan},
  journal   = {Journal of Health Communication},
  pages     = {1--11},
  year      = {2025},
  publisher = {Taylor \& Francis}
}

@article{li2024does,
  title   = {How Does the Disclosure of AI Assistance Affect the Perceptions of Writing?},
  author  = {Li, Zhuoyan and Liang, Chen and Peng, Jing and Yin, Ming},
  journal = {arXiv preprint arXiv:2410.04545},
  year    = {2024}
}

@article{rubin2025comparing,
  title     = {Comparing the value of perceived human versus AI-generated empathy},
  author    = {Rubin, Matan and Li, Joanna Z and Zimmerman, Federico and Ong, Desmond C and Goldenberg, Amit and Perry, Anat},
  journal   = {Nature Human Behaviour},
  pages     = {1--15},
  year      = {2025},
  publisher = {Nature Publishing Group}
}

@article{hwang202580,
  title     = {'It was 80\% me, 20\% AI': Seeking Authenticity in Co-Writing with Large Language Models},
  author    = {Hwang, Angel Hsing-Chi and Liao, Q Vera and Blodgett, Su Lin and Olteanu, Alexandra and Trischler, Adam},
  journal   = {Proceedings of the ACM on Human-Computer Interaction},
  volume    = {9},
  number    = {2},
  pages     = {1--41},
  year      = {2025},
  publisher = {ACM New York, NY, USA}
}

@article{draxler2024ai,
  title     = {The AI ghostwriter effect: When users do not perceive ownership of AI-generated text but self-declare as authors},
  author    = {Draxler, Fiona and Werner, Anna and Lehmann, Florian and Hoppe, Matthias and Schmidt, Albrecht and Buschek, Daniel and Welsch, Robin},
  journal   = {ACM Transactions on Computer-Human Interaction},
  volume    = {31},
  number    = {2},
  pages     = {1--40},
  year      = {2024},
  publisher = {ACM New York, NY}
}

@article{jain2024revealing,
  title     = {Revealing the source: How awareness alters perceptions of AI and human-generated mental health responses},
  author    = {Jain, Gagan and Pareek, Samridhi and Carlbring, Per},
  journal   = {Internet Interventions},
  volume    = {36},
  pages     = {100745},
  year      = {2024},
  publisher = {Elsevier}
}

@article{ovsyannikova2025third,
  title     = {Third-party evaluators perceive AI as more compassionate than expert humans},
  author    = {Ovsyannikova, Dariya and de Mello, Victoria Oldemburgo and Inzlicht, Michael},
  journal   = {Communications Psychology},
  volume    = {3},
  number    = {1},
  pages     = {4},
  year      = {2025},
  publisher = {Nature Publishing Group UK London}
}

@inproceedings{mccann2020fugashi,
    title = "fugashi, a Tool for Tokenizing {J}apanese in Python",
    author = "McCann, Paul",
    editor = "Park, Eunjeong L.  and
      Hagiwara, Masato  and
      Milajevs, Dmitrijs  and
      Liu, Nelson F.  and
      Chauhan, Geeticka  and
      Tan, Liling",
    booktitle = "Proceedings of Second Workshop for NLP Open Source Software (NLP-OSS)",
    month = nov,
    year = "2020",
    address = "Online",
    publisher = "Association for Computational Linguistics",
    url = "https://aclanthology.org/2020.nlposs-1.7/",
    doi = "10.18653/v1/2020.nlposs-1.7",
    pages = "44--51"
}

@article{reimers2020making,
  title={Making monolingual sentence embeddings multilingual using knowledge distillation},
  author={Reimers, Nils and Gurevych, Iryna},
  journal={arXiv preprint arXiv:2004.09813},
  year={2020}
}

@inproceedings{hoque2024hallmark,
  title     = {The HaLLMark effect: Supporting provenance and transparent use of large language models in writing with interactive visualization},
  author    = {Hoque, Md Naimul and Mashiat, Tasfia and Ghai, Bhavya and Shelton, Cecilia D and Chevalier, Fanny and Kraus, Kari and Elmqvist, Niklas},
  booktitle = {Proceedings of the 2024 CHI Conference on Human Factors in Computing Systems},
  pages     = {1--15},
  year      = {2024}
}

@article{yin2024ai,
  title     = {AI can help people feel heard, but an AI label diminishes this impact},
  author    = {Yin, Yidan and Jia, Nan and Wakslak, Cheryl J},
  journal   = {Proceedings of the National Academy of Sciences},
  volume    = {121},
  number    = {14},
  pages     = {e2319112121},
  year      = {2024},
  publisher = {National Academy of Sciences}
}

@article{burleson2010nature,
  title     = {The nature of interpersonal communication},
  author    = {Burleson, Brant R},
  journal   = {The handbook of communication science},
  volume    = {1},
  number    = {2},
  pages     = {145--163},
  year      = {2010},
  publisher = {Sage}
}

@article{schiavo2024comprehension,
  title     = {Comprehension, apprehension, and acceptance: Understanding the influence of literacy and anxiety on acceptance of artificial Intelligence},
  author    = {Schiavo, Gianluca and Businaro, Stefano and Zancanaro, Massimo},
  journal   = {Technology in Society},
  volume    = {77},
  pages     = {102537},
  year      = {2024},
  publisher = {Elsevier}
}

@article{noy2023experimental,
  title     = {Experimental evidence on the productivity effects of generative artificial intelligence},
  author    = {Noy, Shakked and Zhang, Whitney},
  journal   = {Science},
  volume    = {381},
  number    = {6654},
  pages     = {187--192},
  year      = {2023},
  publisher = {American Association for the Advancement of Science}
}

@article{ningrum2023chatgpt,
  title   = {ChatGPT’s impact: The AI revolution in EFL writing},
  author  = {Ningrum, Sulistya and others},
  journal = {Borneo Engineering \& Advanced Multidisciplinary International Journal},
  volume  = {2},
  number  = {Special Issue (TECHON 2023)},
  pages   = {32--37},
  year    = {2023}
}

@article{mahmud2024decoding,
  title     = {Decoding algorithm appreciation: Unveiling the impact of familiarity with algorithms, tasks, and algorithm performance},
  author    = {Mahmud, Hasan and Islam, AKM Najmul and Luo, Xin Robert and Mikalef, Patrick},
  journal   = {Decision Support Systems},
  volume    = {179},
  pages     = {114168},
  year      = {2024},
  publisher = {Elsevier}
}

@article{majovsky2024perfect,
  title     = {Perfect detection of computer-generated text faces fundamental challenges},
  author    = {M{\'a}jovsk{\`y}, Martin and {\v{C}}ern{\`y}, Martin and Netuka, David and Mikolov, Tom{\'a}{\v{s}}},
  journal   = {Cell Reports Physical Science},
  volume    = {5},
  number    = {1},
  year      = {2024},
  publisher = {Elsevier}
}

@article{weber2023testing,
  title     = {Testing of detection tools for AI-generated text},
  author    = {Weber-Wulff, Debora and Anohina-Naumeca, Alla and Bjelobaba, Sonja and Folt{\`y}nek, Tom{\'a}{\v{s}} and Guerrero-Dib, Jean and Popoola, Olumide and {\v{S}}igut, Petr and Waddington, Lorna},
  journal   = {International Journal for Educational Integrity},
  volume    = {19},
  number    = {1},
  pages     = {1--39},
  year      = {2023},
  publisher = {Springer}
}

@article{ayers2023comparing,
  title     = {Comparing physician and artificial intelligence chatbot responses to patient questions posted to a public social media forum},
  author    = {Ayers, John W and Poliak, Adam and Dredze, Mark and Leas, Eric C and Zhu, Zechariah and Kelley, Jessica B and Faix, Dennis J and Goodman, Aaron M and Longhurst, Christopher A and Hogarth, Michael and others},
  journal   = {JAMA internal medicine},
  volume    = {183},
  number    = {6},
  pages     = {589--596},
  year      = {2023},
  publisher = {American Medical Association}
}

@article{proksch2024impact,
  title={The impact of text topic and assumed human vs. AI authorship on competence and quality assessment},
  author={Proksch, Sebastian and Sch{\"u}hle, Julia and Streeb, Elisabeth and Weymann, Finn and Luther, Teresa and Kimmerle, Joachim},
  journal={Frontiers in Artificial Intelligence},
  volume={7},
  pages={1412710},
  year={2024},
  publisher={Frontiers Media SA}
}

@article{lermann2024effects,
  title={The Effects of Assumed AI vs. Human Authorship on the Perception of a GPT-generated Text},
  author={Lermann Henestrosa, Angelica and Kimmerle, Joachim},
  journal={Journalism and Media},
  volume={5},
  number={3},
  pages={1085--1097},
  year={2024},
  publisher={MDPI}
}

@article{porter2024ai,
  title={AI-generated poetry is indistinguishable from human-written poetry and is rated more favorably},
  author={Porter, Brian and Machery, Edouard},
  journal={Scientific Reports},
  volume={14},
  number={1},
  pages={26133},
  year={2024},
  publisher={Nature Publishing Group UK London}
}

@article{kirk2025ai,
  title     = {The AI-authorship effect: Understanding authenticity, moral disgust, and consumer responses to AI-generated marketing communications},
  author    = {Kirk, Colleen P and Givi, Julian},
  journal   = {Journal of Business Research},
  volume    = {186},
  pages     = {114984},
  year      = {2025},
  publisher = {Elsevier}
}

@inproceedings{wang2025intentprism,
  title={IntentPrism: Human-AI Intent Manifestation for Web Information Foraging},
  author={Wang, Zehuan and Xiao, Jiaqi and Sun, Jingwei and Liu, Can},
  booktitle={Proceedings of the Extended Abstracts of the CHI Conference on Human Factors in Computing Systems},
  pages={1--11},
  year={2025}
}

@inproceedings{peng2025navigating,
  title={Navigating the Unknown: A Chat-Based Collaborative Interface for Personalized Exploratory Tasks},
  author={Peng, Yingzhe and Qin, Xiaoting and Zhang, Zhiyang and Zhang, Jue and Lin, Qingwei and Yang, Xu and Zhang, Dongmei and Rajmohan, Saravan and Zhang, Qi},
  booktitle={Proceedings of the 30th International Conference on Intelligent User Interfaces},
  pages={1048--1063},
  year={2025}
}

@inproceedings{he2025contributions,
  title={Which contributions deserve credit? perceptions of attribution in human-ai co-creation},
  author={He, Jessica and Houde, Stephanie and Weisz, Justin D},
  booktitle={Proceedings of the 2025 CHI Conference on Human Factors in Computing Systems},
  pages={1--18},
  year={2025}
}

@article{buchanan2024people,
  title={Do people trust humans more than ChatGPT?},
  author={Buchanan, Joy and Hickman, William},
  journal={Journal of Behavioral and Experimental Economics},
  volume={112},
  pages={102239},
  year={2024},
  publisher={Elsevier}
}

@article{zhang2023human,
  title={Human favoritism, not AI aversion: People’s perceptions (and bias) toward generative AI, human experts, and human--GAI collaboration in persuasive content generation},
  author={Zhang, Yunhao and Gosline, Ren{\'e}e},
  journal={Judgment and Decision Making},
  volume={18},
  pages={e41},
  year={2023},
  publisher={Cambridge University Press}
}

@article{bohm2023people,
  title={People devalue generative AI’s competence but not its advice in addressing societal and personal challenges},
  author={B{\"o}hm, Robert and J{\"o}rling, Moritz and Reiter, Leonhard and Fuchs, Christoph},
  journal={Communications Psychology},
  volume={1},
  number={1},
  pages={32},
  year={2023},
  publisher={Nature Publishing Group UK London}
}

@article{bigman2018people,
  title={People are averse to machines making moral decisions},
  author={Bigman, Yochanan E and Gray, Kurt},
  journal={Cognition},
  volume={181},
  pages={21--34},
  year={2018},
  publisher={Elsevier}
}

@article{morewedge2022preference,
  title={Preference for human, not algorithm aversion},
  author={Morewedge, Carey K},
  journal={Trends in Cognitive Sciences},
  volume={26},
  number={10},
  pages={824--826},
  year={2022},
  publisher={Elsevier}
}

@article{castelo2019task,
  title={Task-dependent algorithm aversion},
  author={Castelo, Noah and Bos, Maarten W and Lehmann, Donald R},
  journal={Journal of marketing research},
  volume={56},
  number={5},
  pages={809--825},
  year={2019},
  publisher={SAGE Publications Sage CA: Los Angeles, CA}
}

@article{dietvorst2015algorithm,
  title={Algorithm aversion: people erroneously avoid algorithms after seeing them err.},
  author={Dietvorst, Berkeley J and Simmons, Joseph P and Massey, Cade},
  journal={Journal of experimental psychology: General},
  volume={144},
  number={1},
  pages={114},
  year={2015},
  publisher={American Psychological Association}
}

@inproceedings{rae2024effects,
  title={The effects of perceived AI use on content perceptions},
  author={Rae, Irene},
  booktitle={Proceedings of the 2024 CHI Conference on Human Factors in Computing Systems},
  pages={1--14},
  year={2024}
}

@article{purcell2024people,
  title={People have different expectations for their own versus others' use of AI-mediated communication tools},
  author={Purcell, Zoe A and Dong, Mengchen and Nussberger, Anne-Marie and K{\"o}bis, Nils and Jakesch, Maurice},
  journal={British Journal of Psychology},
  year={2024},
  publisher={Wiley Online Library}
}

@article{radivojevic2024human,
  title={Human perception of llm-generated text content in social media environments},
  author={Radivojevic, Kristina and Chou, Matthew and Badillo-Urquiola, Karla and Brenner, Paul},
  journal={arXiv preprint arXiv:2409.06653},
  year={2024}
}

@misc{ACM2025AuthorshipPolicy,
author = {{Association for Computing Machinery (ACM)}},
title = {ACM Policy on Authorship},
howpublished = {\url{https://www.acm.org/publications/policies/new-acm-policy-on-authorship}},
note = {Updated on September 16, 2025},
year = {2025},
urldate = {2026-01-06}
}

@article{hwang2023chatgpt,
  title={Is ChatGPT a “Fire of Prometheus” for non-native English-speaking researchers in academic writing?},
  author={Hwang, Sung Il and Lim, Joon Seo and Lee, Ro Woon and Matsui, Yusuke and Iguchi, Toshihiro and Hiraki, Takao and Ahn, Hyungwoo},
  journal={Korean Journal of Radiology},
  volume={24},
  number={10},
  pages={952},
  year={2023}
}

@article{li2024exploring,
  title={Exploring the potential of artificial intelligence to enhance the writing of English academic papers by non-native English-speaking medical students-the educational application of ChatGPT},
  author={Li, Jiakun and Zong, Hui and Wu, Erman and Wu, Rongrong and Peng, Zhufeng and Zhao, Jing and Yang, Lu and Xie, Hong and Shen, Bairong},
  journal={BMC Medical Education},
  volume={24},
  number={1},
  pages={736},
  year={2024},
  publisher={Springer}
}
\clearpage
\appendix
\onecolumn

\section{Statements in the Author Evaluation and Post-experimental Questionnare}

\begin{table}[H]
\small
\centering
\caption{Author's impression evaluation statements used on pre/post-disclosure}
\Description{Author's impression evaluation statements used on pre/post-disclosure}
\begin{tabular}{l}
\toprule
\textbf{Competence}\\
\hline
The author is intelligent / unintelligent.*\\
The author is untrained / trained.\\
The author is inexpert / expert.\\
The author is informed / uninformed.*\\
The author is incompetent / competent.\\
The author is bright / stupid.*\\
\hline
\hline
\textbf{Caring / Goodwill}\\
\hline
The author cares about me / doesn't care about me.*\\
The author has my interests at heart. / doesn't have my interests at heart.*\\
The author is self-centered / not self-centered.\\
The author is concerned with me / unconcerned with me.*\\
The author is insensitive / sensitive.\\
The author is not understanding / understanding.\\
\hline
\hline
\textbf{Trustworthiness}\\
\hline
The author is honest / dishonest*.\\
The author is untrustworthy / trustworthy.\\
The author is honorable / dishonorable*.\\
The author is moral / immoral*.\\
The author is unethical / ethical.\\
The author is phoney / genuine.\\
\hline
\hline
\textbf{Likability}\\
\hline
This writer is friendly.\\
This writer is likable.\\
This writer is warm.\\
This writer is approachable.\\
I would ask this writer for advice.\\
I would like to be friends with this writer.\\
This writer is knowledgeable.\\
\hline
\hline
\textbf{Desire for Future Interaction}\\
\hline
How much would you like to spend time with this writer again in the future?\\
If there were opportunities to interact again with this writer, how likely is it\\
\quad that the two of you could become friends? \\
 \bottomrule
\end{tabular}
\begin{flushleft}
\small
\textit{Note.} * indicates reverse-coded items. Scale ranges from -3 (very negative) to +3 (very positive).
\end{flushleft}
\label{tab:author-impression}
\end{table}

\begin{table}[H]
\small
\centering
\caption{AI literacy assessment used on post-experiment questionnaire}
\Description{Participant's AI Literacy Assessment used on post-experiment questionnaire}
\begin{tabular}{l}
\toprule
\textbf{Apply AI Literacy}\\
\hline
I can operate AI applications in everyday life.\\
I can use AI applications to make my everyday life easier. \\
I can use artificial intelligence meaningfully to achieve my everyday goals. \\
In everyday life, I can use AI in a way that makes my tasks easier. \\
In everyday life, I can work together gainfully with an artificial intelligence. \\
In everyday life, I can communicate gainfully with artificial intelligence. \\
\hline
\hline
\textbf{Detect AI Literacy}\\
\hline
I can tell if I am dealing with an application based on artificial intelligence. \\
I can distinguish devices that use AI from devices that do not. \\
I can distinguish if I interact with an AI or a 'real human'. \\
\hline
\hline
\textbf{AI Persuasion Literacy}\\
\hline
I don't let AI influence me in my everyday decisions. \\
I can prevent an AI from influencing me in my everyday decisions. \\
I realise if artificial intelligence is influencing me in my everyday decisions. \\
\bottomrule
\end{tabular}
\begin{flushleft}
\small
\textit{Note.} Scale ranges from -3 (strongly disagree) to +3 (strongly agree).
\end{flushleft}
\label{tab:literacy}
\end{table}

\section{Detailed Result of Regression Model}

% order: trust, caring, competence, likability, future
\begin{table*}[h]
  \small
  \centering
  \caption{Mixed Linear Model Regression Results for \textit{TrustworthinessShift}. Marginal R²: 0.2173, Conditional R²: 0.4512}
  \Description{Mixed linear model regression table for \textit{TrustworthinessShift} with Act type (6 categories, reference: To Describe), AIRatio, and AI literacy measures as predictors. Shows coefficients, standard errors, z-values, p-values, and 95\% confidence intervals. Significant main effects include AIRatio (negative) and To Interact act type (negative). Significant positive interaction between AIRatio and Apply AI Literacy.}
  \begin{tabular}{lrrrlcc}
  \toprule
  Variable & Coef. & SE & z & P>|z| & \multicolumn{2}{c}{95\%CI} \\
  \midrule
  \multicolumn{7}{l}{\textbf{Main Effect}} \\
  \textbf{Intercept} & -4.676 & 0.511 & -9.158 & \textbf{0.000***} & -5.677 & -3.675 \\
  \multicolumn{7}{l}{Act (Ref: To Describe)} \\
  \quad To Convince & 0.085 & 0.644 & 0.132 & 0.895 & -1.178 & 1.348 \\
  \quad To Explore & 0.536 & 0.646 & 0.830 & 0.406 & -0.729 & 1.802 \\
  \quad To Imagine & 1.080 & 0.650 & 1.662 & 0.097 & -0.194 & 2.354 \\
  \quad \textbf{To Interact} & -1.999 & 0.650 & -3.076 & \textbf{0.002**} & -3.273 & -0.726 \\
  \quad To Reflect & -0.388 & 0.650 & -0.596 & 0.551 & -1.662 & 0.887 \\
  \textbf{AIRatio} & -3.236 & 0.192 & -16.858 & \textbf{0.000***} & -3.612 & -2.860 \\
  Apply AI Literacy & 0.523 & 0.313 & 1.667 & 0.095 & -0.092 & 1.137 \\
  Detect AI Literacy & 0.473 & 0.312 & 1.516 & 0.130 & -0.139 & 1.084 \\
  \addlinespace
  \multicolumn{7}{l}{\textbf{Interaction}} \\
  \textbf{AIRatio : Apply AI Literacy} & 0.630 & 0.191 & 3.307 & \textbf{0.001**} & 0.257 & 1.004 \\
  \midrule
  Group Var & 13.714 & 0.400 & & & & \\
  \bottomrule
  \multicolumn{7}{l}{\footnotesize * p < 0.05, ** p < 0.01, *** p < 0.001}
  \end{tabular}
  \label{tab:trustworthiness_mlm}
\end{table*}

\begin{table*}[h]
  \small
  \centering
  \caption{Mixed Linear Model Regression Results for \textit{CaringShift}. Marginal R²: 2.036, Conditional R²: 0.3152}
  \begin{tabular}{lrrrlcc}
  \toprule
  Variable & Coef. & SE & z & P>|z| & \multicolumn{2}{c}{95\%CI} \\
  \midrule
  \multicolumn{7}{l}{\textbf{Main Effect}} \\
  \textbf{Intercept} & -2.446 & 0.459 & -5.329 & \textbf{0.000***} & -3.345 & -1.546 \\
  \multicolumn{7}{l}{Act (Ref: To Describe)} \\
  \quad To Convince & -1.075 & 0.619 & -1.737 & 0.082 & -2.287 & 0.138 \\
  \quad \textbf{To Explore} & 1.612 & 0.620 & 2.601 & \textbf{0.009**} & 0.397 & 2.826 \\
  \quad To Imagine & 0.410 & 0.622 & 0.659 & 0.510 & -0.809 & 1.628 \\
  \quad \textbf{To Interact} & -3.050 & 0.622 & -4.903 & \textbf{0.000***} & -4.269 & -1.831 \\
  \quad To Reflect & -0.144 & 0.622 & -0.232 & 0.817 & -1.363 & 1.075 \\
  \textbf{AIRatio} & -2.838 & 0.454 & -6.251 & \textbf{0.000***} & -3.727 & -1.948 \\
  Detect AI Literacy & 0.511 & 0.265 & 1.927 & 0.054 & -0.009 & 1.031 \\
  AI Persuasion Literacy & -0.001 & 0.266 & -0.003 & 0.997 & -0.522 & 0.521 \\
  \addlinespace
  \multicolumn{7}{l}{\textbf{Interaction}} \\
  \multicolumn{7}{l}{Interaction between AIRatio and Act (Ref: To Describe)} \\
  \quad AIRatio : To Convince & -0.390 & 0.643 & -0.606 & 0.544 & -1.651 & 0.871 \\
  \quad AIRatio : To Explore & 0.892 & 0.645 & 1.383 & 0.167 & -0.372 & 2.157 \\
  \quad AIRatio : \textbf{To Imagine} & 1.372 & 0.644 & 2.129 & \textbf{0.033*} & 0.109 & 2.635 \\
  \quad AIRatio : To Interact & -0.508 & 0.641 & -0.792 & 0.428 & -1.765 & 0.749 \\
  \quad AIRatio : To Reflect & 1.098 & 0.646 & 1.700 & 0.089 & -0.168 & 2.365 \\
  \textbf{AIRatio: AI Persuasion Literacy} & 0.467 & 0.184 & 2.538 & \textbf{0.011*} & 0.106 & 0.828 \\
  \midrule
  Group Var & 4.951 & 0.241 & & & & \\
  \bottomrule
  \multicolumn{7}{l}{\footnotesize * p < 0.05, ** p < 0.01, *** p < 0.001}
  \end{tabular}
  \label{tab:caring_mlm}
\end{table*}

\begin{table*}[h]
  \centering
  \caption{Mixed Linear Model Regression Results for \textit{CompetenceShift}. Marginal R²: 0.1385, Conditional R²: 0.3337}
  \Description{Mixed linear model regression table for \textit{CompetenceShift} with Act type (6 categories, reference: To Describe), AIRatio, and AI literacy measures as predictors. Shows coefficients, standard errors, z-values, p-values, and 95\% confidence intervals. Significant main effects include AIRatio (negative), To Convince act type (positive), To Imagine act type (positive), and To Interact act type (positive). Significant positive interaction between AIRatio and Apply AI Literacy.}
  \begin{tabular}{lrrrlcc}
  \toprule
  Variable & Coef. & SE & z & P>|z| & \multicolumn{2}{c}{95\%CI} \\
  \midrule
  \multicolumn{7}{l}{\textbf{Main Effect}} \\
  \textbf{Intercept} & -3.866 & 0.385 & -10.040 & \textbf{0.000***} & -4.620 & -3.111 \\
  \multicolumn{7}{l}{Act (Ref: To Describe)} \\
  \quad\textbf{To Convince} & 1.496 & 0.502 & 2.984 & \textbf{0.003**} & 0.513 & 2.479 \\
  \quad To Explore & 0.247 & 0.503 & 0.491 & 0.624 & -0.739 & 1.233 \\
  \quad \textbf{To Imagine} & 1.335 & 0.506 & 2.637 & \textbf{0.008**} & 0.343 & 2.328 \\
  \quad \textbf{To Interact} & 1.853 & 0.506 & 3.662 & \textbf{0.000***} & 0.861 & 2.845 \\
  \quad To Reflect & 0.801 & 0.507 & 1.580 & 0.114 & -0.193 & 1.794 \\
  \textbf{AIRatio} & -1.774 & 0.149 & -11.913 & \textbf{0.000***} & -2.066 & -1.482 \\
  Apply AI Literacy & 0.220 & 0.221 & 0.997 & 0.319 & -0.213 & 0.653 \\
  Detect AI Literacy & 0.361 & 0.219 & 1.648 & 0.099 & -0.068 & 0.791 \\
  \addlinespace
  \multicolumn{7}{l}{\textbf{Interaction}} \\
  \textbf{AIRatio : Apply AI Literacy} & 0.478 & 0.151 & 3.153 & \textbf{0.002**} & 0.181 & 0.775 \\
  \midrule
  Group Var & 5.763 & 0.253 & & & & \\
  \bottomrule
  \multicolumn{7}{l}{\footnotesize * p < 0.05, ** p < 0.01, *** p < 0.001}
  \end{tabular}
  \label{tab:competency_mlm}
\end{table*}

\begin{table*}[h]
  \small
  \centering
  \caption{Mixed Linear Model Regression Results for \textit{LikabilityShift}. Marginal R²: 0.2227, Conditional R²: 0.2920}
  \Description{Mixed linear model regression table for \textit{LikabilityShift} with Act type (6 categories, reference: To Describe), AIRatio, and AI literacy measures as predictors. Shows coefficients, standard errors, z-values, p-values, and 95\% confidence intervals. Significant main effects include AIRatio (negative), To Interact act type (negative), and Detect AI Literacy (positive). Significant positive interaction between AIRatio and Detect AI Literacy.}
  \begin{tabular}{lrrrlcc}
  \toprule
  Variable & Coef. & SE & z & P>|z| & \multicolumn{2}{c}{95\%CI} \\
  \midrule
  \multicolumn{7}{l}{\textbf{Main Effect}} \\
  \textbf{Intercept} & -4.013 & 0.613 & -6.546 & \textbf{0.000***} & -5.215 & -2.811 \\
  \multicolumn{7}{l}{Act (Ref: To Describe)} \\
  \quad To Convince & 0.206 & 0.841 & 0.245 & 0.806 & -1.442 & 1.855 \\
  \quad To Explore & 1.384 & 0.842 & 1.643 & 0.100 & -0.267 & 3.034 \\
  \quad To Imagine & -0.090 & 0.844 & -0.106 & 0.915 & -1.744 & 1.565 \\
  \quad \textbf{To Interact} & -3.800 & 0.846 & -4.494 & \textbf{0.000***} & -5.457 & -2.143 \\
  \quad To Reflect & -1.094 & 0.845 & -1.295 & 0.195 & -2.750 & 0.562 \\
  \textbf{AIRatio} & -3.729 & 0.246 & -15.148 & \textbf{0.000***} & -4.212 & -3.247 \\
  \textbf{Detect AI Literacy} & 0.854 & 0.281 & 3.039 & \textbf{0.002**} & 0.303 & 1.404 \\
  \addlinespace
  \multicolumn{7}{l}{\textbf{Interaction}} \\
  \textbf{AIRatio : Detect AI Literacy} & 0.731 & 0.246 & 2.976 & \textbf{0.003**} & 0.250 & 1.212 \\
  \midrule
  Group Var & 5.550 & 0.237 & & & & \\
  \bottomrule
  \multicolumn{7}{l}{\footnotesize * p < 0.05, ** p < 0.01, *** p < 0.001}
  \end{tabular}
  \label{tab:likability_mlm}
\end{table*}

\begin{table*}[h]
  \small
  \centering
  \caption{Mixed Linear Model Regression Results for \textit{FutureShift}. Marginal R²: 0.1435, Conditional R²: 0.2245}
  \Description{Mixed linear model regression table for \textit{FutureShift} with AIRatio and AI literacy measures as predictors. Shows coefficients, standard errors, z-values, p-values, and 95\% confidence intervals. Significant main effect of AIRatio (negative). Significant positive interaction between AIRatio and AI Persuasion Literacy.}
  \begin{tabular}{lrrrlcc}
  \toprule
  Variable & Coef. & SE & z & P>|z| & \multicolumn{2}{c}{95\%CI} \\
  \midrule
  \multicolumn{7}{l}{\textbf{Main Effect}} \\
  \textbf{Intercept} & -1.074 & 0.094 & -11.479 & \textbf{0.000***} & -1.257 & -0.890 \\
  \textbf{AIRatio} & -1.034 & 0.081 & -12.705 & \textbf{0.000***} & -1.194 & -0.875 \\
  AI Persuasion Literacy & 0.105 & 0.094 & 1.123 & 0.261 & -0.078 & 0.289 \\
  \addlinespace
  \multicolumn{7}{l}{\textbf{Interaction}} \\
  \textbf{AIRatio: AI Persuasion Literacy} & 0.224 & 0.082 & 2.738 & \textbf{0.006**} & 0.064 & 0.384 \\
  \midrule
  Group Var & 0.645 & 0.093 & & & & \\
  \bottomrule
  \multicolumn{7}{l}{\footnotesize * p < 0.05, ** p < 0.01, *** p < 0.001}
  \end{tabular}
  \label{tab:future_mlm}
\end{table*}

\end{document}